\documentclass[onecolumn,secnumarabic,nobibnotes,superscriptaddress,aps]{revtex4-2}
\usepackage{color, xcolor, colortbl}
\usepackage{graphicx}
\usepackage{geometry}
\usepackage{amsthm,amsmath,amssymb,amsfonts}
\usepackage{dcolumn}
\usepackage{url}
\usepackage{epstopdf}
\usepackage{enumitem}
\usepackage{algorithm}
\usepackage{algpseudocode}
\usepackage{bm}
\usepackage[caption=false]{subfig}
\usepackage{appendix}
\usepackage{multirow}
\usepackage{braket}
\usepackage[english]{babel}
\usepackage{hyperref}
\usepackage[capitalize]{cleveref}
\usepackage{xpatch}
\usepackage{tikz}
\usepackage{adjustbox}
\usepackage{xspace}
\usetikzlibrary{quantikz}
\newcommand{\bvec}[1]{\mathbf{#1}}

\newcommand{\vb}{\bvec{b}}

\renewcommand{\Re}{\operatorname{Re}}
\renewcommand{\Im}{\operatorname{Im}}

\newcommand{\I}{i}

\newcommand{\mc}[1]{\mathcal{#1}}

\newcommand{\wt}[1]{\widetilde{#1}}

\newcommand{\abs}[1]{\left\lvert#1\right\rvert}
\newcommand{\norm}[1]{\left\lVert#1\right\rVert}

\newcommand{\Or}{\mathcal{O}}

\newcommand{\NN}{\mathbb{N}}
\newcommand{\RR}{\mathbb{R}}
\newcommand{\CC}{\mathbb{C}}

\newtheorem{thm}{\protect\theoremname}
\newtheorem{lem}[thm]{\protect\lemmaname}

\newtheorem{prop}[thm]{\protect\propositionname}
\newtheorem{cor}[thm]{\protect\corollaryname}

\newtheorem{defn}[thm]{\protect\definitionname}

\providecommand{\definitionname}{Definition}
\providecommand{\assumptionname}{Assumption}
\providecommand{\corollaryname}{Corollary}
\providecommand{\lemmaname}{Lemma}
\providecommand{\propositionname}{Proposition}
\providecommand{\remarkname}{Remark}
\providecommand{\theoremname}{Theorem}
\providecommand{\problemname}{Problem}
\usepackage{bbm}

\makeatletter
\newenvironment{breakablealgorithm}
  {
   \begin{center}
     \refstepcounter{algorithm}
     \hrule height.8pt depth0pt \kern2pt
     \renewcommand{\caption}[2][\relax]{
       {\raggedright\textbf{\fname@algorithm~\thealgorithm} ##2\par}%
       \ifx\relax##1\relax 
         \addcontentsline{loa}{algorithm}{\protect\numberline{\thealgorithm}##2}%
       \else 
         \addcontentsline{loa}{algorithm}{\protect\numberline{\thealgorithm}##1}%
       \fi
       \kern2pt\hrule\kern2pt
     }
  }{
     \kern2pt\hrule\relax
   \end{center}
  }
\makeatother

\usetikzlibrary{fit}
\tikzset{%
  highlight/.style={rectangle,rounded corners,fill=blue!15,draw,fill opacity=0.3,thick,inner sep=0pt}
}

\newcommand{\DeptMath}{Department of Mathematics, University of California, Berkeley, California 94720 USA\\}
\newcommand{\Google}{Google Quantum AI, Venice, CA 90291, USA}
\newcommand{\umd}{Joint Center for Quantum Information and Computer Science, University of Maryland, College Park, MD 20742, USA}
\newcommand{\ucsb}{University of California, Santa Barbara, California, 93106, USA.}

\usepackage{dsfont}
\newcommand{\bP}{\mathbb{P}}

\newcommand{\tr}[1]{\mathrm{tr}\left(#1\right)}
\newcommand{\ctrll}{\text{ctrl}}

\newcommand{\ffqc}{\text{FQSVT}\xspace}
\newcommand{\offqc}{\text{1-FQSVT}\xspace}

\usepackage[normalem]{ulem}

\begin{document}

\title{Feedforward Quantum Singular Value Transformation }
\author{Yulong Dong}
\affiliation{\DeptMath}
\author{Dong An}
\affiliation{\umd}

\author{Murphy Yuezhen Niu}
\email[Electronic address: ]{murphyniu@ucsb.edu}
\affiliation{\Google}
\affiliation{\ucsb}
\date{\today}

\begin{abstract}
In this paper, we introduce a major advancement in Quantum Singular Value Transformation (QSVT) through the development of Feedforward QSVT (FQSVT), a framework that significantly enhances the efficiency and robustness of quantum algorithm design. By leveraging intermediate measurements and feedforward operations, FQSVTs reclaim quantum information typically discarded in conventional QSVT, enabling more efficient transformations. Our results show that FQSVTs can exponentially accelerate the projection of quantum states onto energy subspaces, outperforming probabilistic projection and adiabatic algorithms with superior efficiency and a drastic reduction in query complexity. In the context of superconducting qubits, FQSVTs offer a powerful tool for managing energy subspaces, improving efficiency for state preparation and leakage detection.
\end{abstract}

\maketitle

\section{Introduction}

Quantum Singular Value Transformation (QSVT) is one of the most fascinating algorithms in modern quantum computing. Efficient quantum algorithms have been demonstrated through QSVT for a variety of problems, including quantum Hamiltonian simulation \cite{LowChuang2017,GilyenSuLowEtAl2019}, ground state preparation \cite{LinTong2020near,DongLinTong2022,DongLin2024}, quantum linear system problems \cite{LinTong2020}, and quantum benchmarking \cite{DongLin2021,DongWhaleyLin2021}. Additionally, the QSVT framework provides a modern, unified approach to understanding and designing quantum algorithms across diverse fields \cite{MartynRossiTanEtAl2021}. Traditionally, QSVT only utilizes part of its dynamics, relying on a single projective measurement of the ancillary system at the end of the process. This approach restricts the types of transformations QSVT can achieve and limits the overall success probability of reaching the desired transformation.

Recent discoveries have shown that incorporating intermediate measurement information is essential for extending QSVT to include the Quantum Fourier Transform algorithm~\cite{MartynRossiTanEtAl2021}. Meanwhile, varying circuit structure adaptive to intermediate measurement outcome also leads to a heuristic quantum algorithm for the preparation of the ground state that emulates thermodynamic cooling \cite{MotlaghZiniArrazola2024}. In addition, conditional resampling based on measurement outcomes is crucial for developing dissipative algorithms for ground-state preparation for local or frustration-free Hamiltonians \cite{VerstraeteWolfCirac2009,Cubitt2023}.  These findings inspire an intriguing question:
\begin{center}
    \textit{Can we develop a  framework for QSVT to incorporate operations adaptive to intermediate measurements, and demonstrate its advantage over other exisiting methods in applications?}
\end{center}

In this paper, we provide a positive answer to this question in one setting. We introduce a crucial ingredient to QSVT that has yet to be studied systematically: intermediate measurements and feedforward controlled operation. In our approach, quantum information that would typically be discarded in conventional QSVT measurements is repurposed for additional transformations. This technique enables us to adjust the overall output to achieve the desired targets more efficiently than algorithms without the feedforward structure. We demonstrate that this new framework allows us to perform specific tasks, such as projecting an unknown state to an energy subspace, exponentially faster than existing alternative methods.

The paper is structured as follows: In \cref{sec:fqsvt}, we define feedforward QSVT (\ffqc), starting with its construction from conventional QSVT circuits and mid-circuit measurements, and provide a formal definition. Next, in \cref{sec:multi-band-problem}, we introduce the multi-band projection problem. \cref{sec:multi-band-FQSVT} details a quantum algorithm to solve this problem using our \ffqc framework and analyzes its complexity. In \cref{sec:advantage}, we highlight the advantages of feedforward by comparing the complexity of our method with other existing approaches. The application of the multi-band projection problem to superconducting qubits is discussed in \cref{sec:superconducting}. The paper concludes with a discussion in \cref{sec:discussion}.

\section{Quantum singular value transformations with feedforward operations}\label{sec:fqsvt}
\begin{figure}[htbp]
    \centering
    \includegraphics[width=\linewidth]{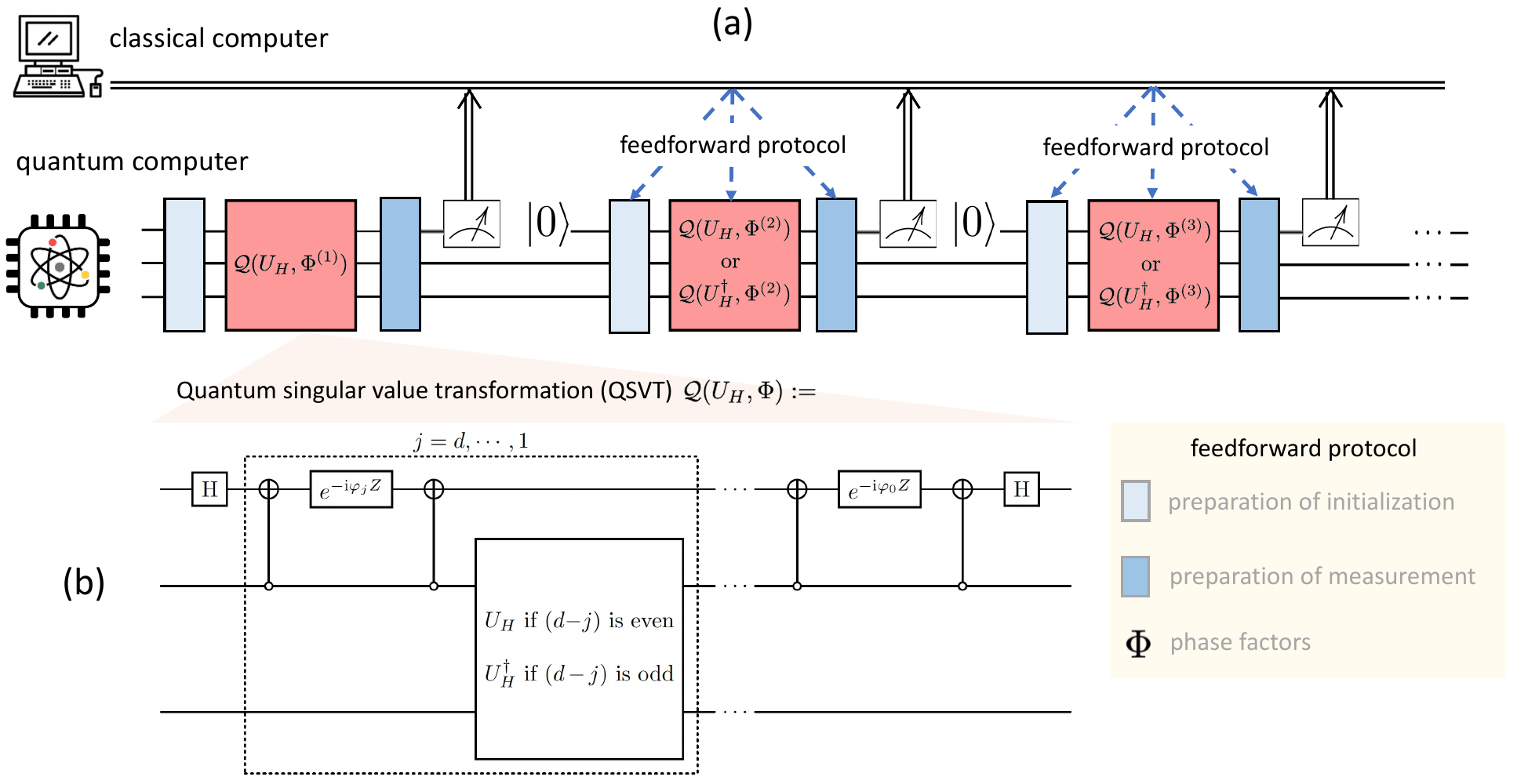}
    \caption{A framework of \ffqc. (a) The circuit structure of \ffqc. (b) The circuit structure of conventional QSVT. Each quantum circuit block contains an initialization, an action of conventional QSVT circuit, and a measurement basis preparation. The measurement information is maintained in a classical computer. It also derives feedforward protocols to construct future quantum circuit blocks based on the record of measurement outcomes. }
    \label{fig:feedforward_mainfig}
\end{figure}

Quantum singular value transformation (QSVT) is an important quantum algorithm primitive for designing efficient quantum algorithms \cite{LowChuang2017,GilyenSuLowEtAl2019}. Moreover, recent research also demonstrates that QSVT provides a unified framework for understanding a variety of quantum algorithms and applications \cite{MartynRossiTanEtAl2021}. The theory of QSVT is built on an approximation theorem in terms of product of parametric $\mathrm{SU}(2)$ matrices, which is also referred to as \textit{Quantum Signal Processing (QSP)}, and a matrix decomposition that lifts QSP to high-dimensional matrices. To accommodate the unitary nature of quantum operations, the relevant matrix $H$ is accessed through an input model called \textit{block encoding}, which is a unitary $U_H$ whose upper-left submatrix is equal to $H$ (up to rescaling) \cite{ChakrabortyGilyenJeffery2018,LowChuang2019,GilyenSuLowEtAl2019}. QSVT interleaves block encodings and a sequence of controlled $Z$-rotations to implement a matrix function of $H$. Since the application of $U_H$ may transform the basis of the original Hilbert space, QSVT compensates for such unwanted basis transformation by interleaving $U_H$ and its inverse $U_H^\dagger$ in the circuit construction, which is detailed in the bottom panel in \cref{fig:feedforward_mainfig}. For simplicity, we assume $H$ is a Hermitian matrix, whose transformed matrix function coincides with the conventional definition of matrix functions through diagonalization. 

Assume the block encoding $U_H$ of $H$ requires the use of extra $m$ ancilla qubits. Taking account the additional single qubit for implementing controlled rotation, the standard application of conventional QSVT is written as an action on the input quantum state, 
\begin{equation}
    \mc{Q}(U_H, \Phi) \ket{0^{m + 1}} \ket{\phi} = \ket{0^{m + 1}} f(H) \ket{\phi} + \ket{\bot},
\end{equation}
where the success of implementing a matrix function $f(H)$ is flagged by an all-zero ancillary qubit state. Here, $\ket{\bot}$ represents a possibly unnormalized state which is orthogonal to the ``successful'' ancillary state such that $\left(\bra{0^{m+1}} \otimes I_N\right) \ket{\bot} = 0$. It is considered as a ``gabage'' state in conventional QSVT applications and it is discarded in classically repeated measurements or is coherently deamplified using amplitude amplification method. However, the ``garbage'' state can contain quantum information that can be further converted towards solving the target problem with feedforward operations. 

To exploit these overlooked quantum information, we propose a framework by equipping QSVT with feedforward operations. To partially probe information in the system, a subset of ancilla qubits is isolated to perform measurement and subsequent reset operation. These qubits are referred to as \textit{monitoring qubits} in our framework. The measurement outcomes of monitoring qubits are stored in classical memory for further use. Note that the quantum state also collapses in accordance with the measurement outcomes. The collapsed quantum state can be the desired state of our interest or lies in the orthogonal complement where further conversion is needed to render it useful. The subsequent quantum circuit construction is based on the record of measurement outcomes. Processing with classical computer, subsequent feedforward operations are instructed in terms of preparations of initialization and measurement and the set of phase factors in the next round. The set of phase factors determines the subsequent matrix function to be applied, which stands for some correction and conversion at the level of matrix functions. Meanwhile, the alternative choice of block encoding and its inverse is to account for potential cancellation of basis transformation of the underlying Hilbert space. As these operations are conditioned on the record of measurement outcomes, they generalize the conventional framework of QSVT to integrate feedforward operations. We also discuss the concept of feedforwarding that is  applicable to a broad range of quantum algorithm structures beyond QSVT  in \cref{sec:general-feedforward}.

\section{Multi-band projection problem and its solution using feedforward}\label{sec:projection}

In this section, we provide a concrete problem that exemplifies the advantage of the use of feedforward QSVT (\ffqc). We aim to prepare a quantum state within a specific energy subspace of the target Hamiltonian, characterized by its energy band structure. We demonstrate that \ffqc can prepare such a state with exponentially less query to the target Hamiltonian than the existing methods.  

\subsection{Problem statement of multi-band projection}\label{sec:multi-band-problem}
We consider an $n$-qubit Hamiltonian $H = \sum_{i \in [N]} E_i \ket{\phi_i}\bra{\phi_i}$. Without loss of generality, we assume the Hamiltonian is positive semidefinite, and its eigenvalues are sorted ascending $0 \le E_0 \le E_1 \le \cdots \le E_{N-1} \le \norm{H}$. The eigenvalues of the Hamiltonian are clustered into $L$ bands, which are referred to as \textit{energy bands}. Equivalently, there are $(L-1)$ subintervals, which are referred to as \textit{energy gaps}, in the energy spectrum so that they contain no eigenvalues of the Hamiltonian. The $j$-th energy gap separates the $j$-th and the $(j+1)$-th energy bands. Suppose the center of the $j$-th gap is $\mu_j$ and the (minimal) spread among gaps is $\Delta$. The energy band assumption means:
\begin{equation}
    \mathrm{eig}(H) \cap \left(\bigcup_{j = 0}^{L - 2}(\mu_j - \Delta/2, \mu_j + \Delta/2)\right) = \emptyset \text{ with } \mathrm{eig}(H) := \{E_i : i \in [N]\}.
\end{equation}

These gaps separate the energy spectrum into $L$ nonoverlapping bands. The first band and the last bands contain energy indices $\mc{B}_0 = \{i : E_i < \mu_0\}$ and $\mc{B}_{L-1} = \{i : E_i \ge \mu_{L - 2}\}$ respectively. Other energy bands belong to energy indices $\mc{B}_j = \{i : \mu_{j - 1} \le E_i < \mu_j\}$. The projection onto the energy subspace of the $j$-th energy band is represented by: 
\begin{equation}
    \Pi_j = \sum_{i \in \mc{B}_j} \ket{\phi_i} \bra{\phi_i},
\end{equation}
which projects an arbitrary quantum state onto a subspace within the given energy band. The preparation of such band-supported quantum state has applications in estimating density of states \cite{DongLin2021},  studying excited-state chemistry and physics~\cite{excited2,excited3} and error mitigation~\cite{kandala2019error}.  The goal of solving the multi-band projection problem is formalized as implementing a quantum channel 
\begin{equation}
    \mc{E}_\mathrm{proj}(\varrho) = \sum_{j \in [L]} \Pi_j \varrho \Pi_j
\end{equation}
which fully separates the input quantum state in terms of energy bands. 

\subsection{Solving multi-band projection problem using feedforward QSVT}\label{sec:multi-band-FQSVT}
As discussed in the previous section, conventional QSVT applications introduce a ``garbage'' state which is usually discarded in practice. However, by performing detailed analyses of the circuit structure in \cref{cor:bot-state}, we demonstrate that such conventionally considered ``gabage'' state contains meaningful quantum information which can be converted to useful content with some subsequent operations. By choosing a set of \textit{symmetric} QSVT phase factors \cite{DongMengWhaleyEtAl2021,WangDongLin2021}, measuring a single ancilla suffices to indicate the ``successful'' application of QSVT. This qubit, visualized as the top ancilla in the QSVT circuit (see bottom panel of \cref{fig:feedforward_mainfig}), is referred to as the \textit{monitoring qubit}. Hence, the choice of subsequent operations is conditioned on the measurement outcome of the monitoring qubit. These conditional operations lead to the first technical result conceptualizing our \ffqc circuit structure. As the feedforward operation only depends on the previous measurement result, the circuit structure is referred to as a one-step feedforward quantum singular value transformation circuit (\offqc). 

\begin{figure*}[htbp]
\begin{center}
\subfloat[]{
\includegraphics[width=\textwidth]{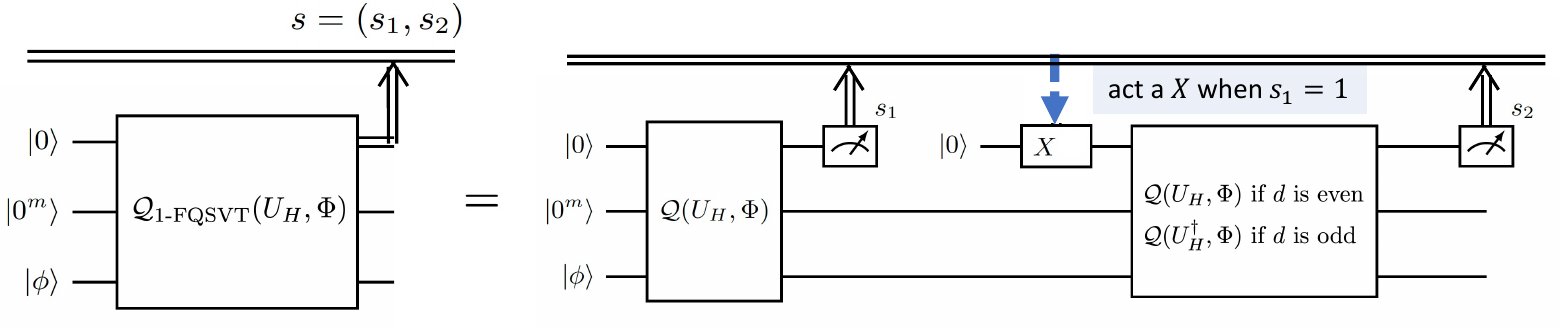}
}
\hfill
\subfloat[]{
\includegraphics[width=\textwidth]{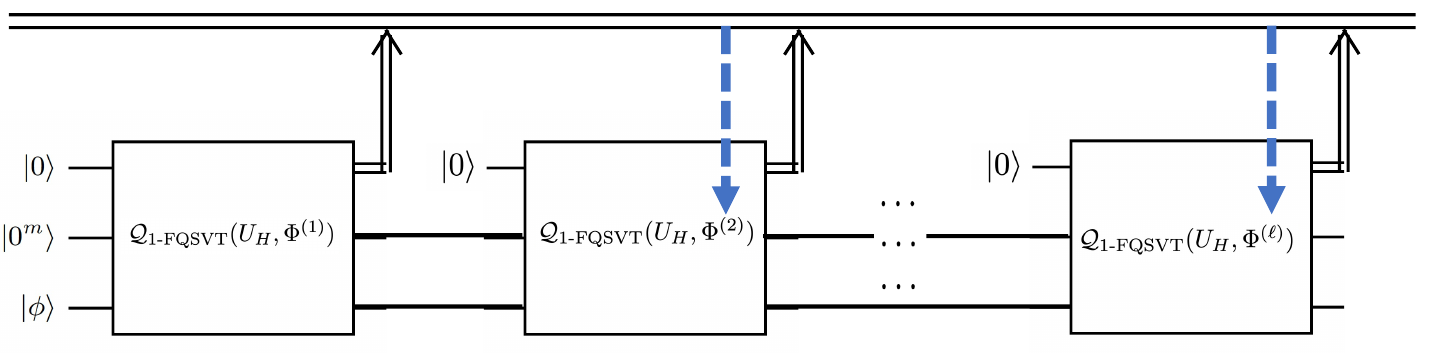}
}
\end{center}
\caption{\ffqc circuits for solving multi-band projection problem. (a) \offqc circuit that implements a binary projection. The feedforward protocol applies a Pauli $X$ gate as initialization of the second circuit block according to the measurement outcome $s_1$. (b) \ffqc by stacking multiple \offqc circuits. The measurement outcomes are maintained and are used to control future QSVT phase factors. Remarkably, this controlled-QSVT circuit is implemented by simply controlling those single-qubit $Z$-rotation gates acting on the top monitoring ancillary qubit (see \cref{fig:feedforward_mainfig} (b)) The circuit is constructed according to \cref{alg:feedforward_projector}.}
\label{fig:1FF-QSVT-construction}
\end{figure*}

We explicitly give the construction of the \offqc circuit in \cref{fig:1FF-QSVT-construction}, which utilizes two conventional QSVT circuits while the second circuit's construction depends on the measurement outcome of the moniroting qubits of the previous one. Such conditional feedforward operations serve as the core of converting the ``failure'' of the first QSVT circuit to useful quantum information. This is conceptualized as the following technical theorem, whose proof can be found in \cref{sec:proof-FFQC-projection}.
\begin{thm}\label{lma:1FF-QSVT}
    Let $\Phi$ be a set of symmetric phase factors implementing a real polynomial $f \in \RR[x]$. Let $U_H$ be a unitary block encoding of a Hamiltonian $H$ with bounded norm $\norm{H} \le 1$, and $\ket{\phi}$ be an $n$-qubit input state. Consider a set of measurement outcomes of the monitoring qubit $s := (s_1, s_2)$ of a \offqc in \cref{fig:1FF-QSVT-construction}. When $s = (0, 0)$, the final unnormalized quantum state is $\ket{\phi;s} = f^2(H)\ket{\phi}$. When $s = (1, 0)$, the final unnormalized quantum state is $\ket{\phi;s} = -\left(1-f^2(H)\right)\ket{\phi}$. The probability of measuring $s$ is $p_s = \norm{\ket{\phi;s}}^2$.
\end{thm}

An important application of QSVT is implementing projectors based on an energy threshold value $\mu$ of a given Hamiltonian. Namely, we would like to implement a projector $\Pi_{<\mu}$ such that $\Pi_{<\mu}\ket{\phi_i} = \mathbf{1}_{E_i < \mu} \ket{\phi_i}$ for any eigenstate $\ket{\phi_i}$ of $H$.  This task is conceptualized as implementing an even polynomial approximating Heaviside function, which satisfies the following conditions:
\begin{align}
    &\abs{f(x)} < \epsilon/2 \quad \forall x > \mu + \Delta / 2, \label{eqn:filter1}\\
    &\abs{1-f(x)} < \epsilon/2 \quad \forall 0 \le x < \mu-\Delta/2, \label{eqn:filter2}\\
    &\abs{f(x)} \le 1 \quad \forall -1 \le x \le 1.\label{eqn:filter3}
\end{align}
Note that the ambiguity within a transition gap of width $\Delta$ is to compensate for the discontinuity of the Heaviside function. Consequently, when there is no energy eigenvalue lying in the transition gap, the matrix function $f(H)$ approximates the projector onto the subspace whose energy is lower than the threshold value $\mu$. Previous work \cite{WanBertaCampbell2021randomized} shows that a QSVT circuit, which approximately implements that projector, can be derived using $d = \Or\left(\norm{H} \Delta^{-1}\log(\epsilon^{-1})\right)$ queries to the block encoding $U_H$ and its inverse. In conventional applications, when a supplementary projector is needed to project the quantum state onto the high-energy subspace, a separate QSVT circuit implementing \(1 - f\) must be derived due to the absence of a feedforward mechanism. This increases the query cost for QSVT to implement both projectors due to the use of additional quantum circuits and the need to amplify the ``successful'' component. In contrast, by introducing the feedforward mechanism, according to \cref{lma:1FF-QSVT}, both projectors can be implemented simultaneously using QSVT with feedforward mechanism. 

\begin{cor}[\offqc for projection]\label{cor:1FF-QSVT-projection}
    When $f$ is a polynomial approximation to the Heaviside function satisfying \cref{eqn:filter1,eqn:filter2,eqn:filter3} and $\mathrm{eig}(H) \cap (\mu-\Delta/2,\mu+\Delta/2) = \emptyset$, the \offqc circuit satisfies \textit{(1)} $\norm{\Pi_{< \mu}\ket{\phi} - \ket{\phi;(0,0)}} < \epsilon$, \textit{(2)} $\norm{-\Pi_{\ge \mu}\ket{\phi} - \ket{\phi;(1,0)}} < \epsilon$, and \textit{(3)} the probability of measuring other outcomes is small $\bP(s = (0,1)\text{ or }(1,1)) \le 2\sqrt{2} \epsilon$.
\end{cor}

\begin{figure}[htbp]
    \centering
    \includegraphics[width = \textwidth]{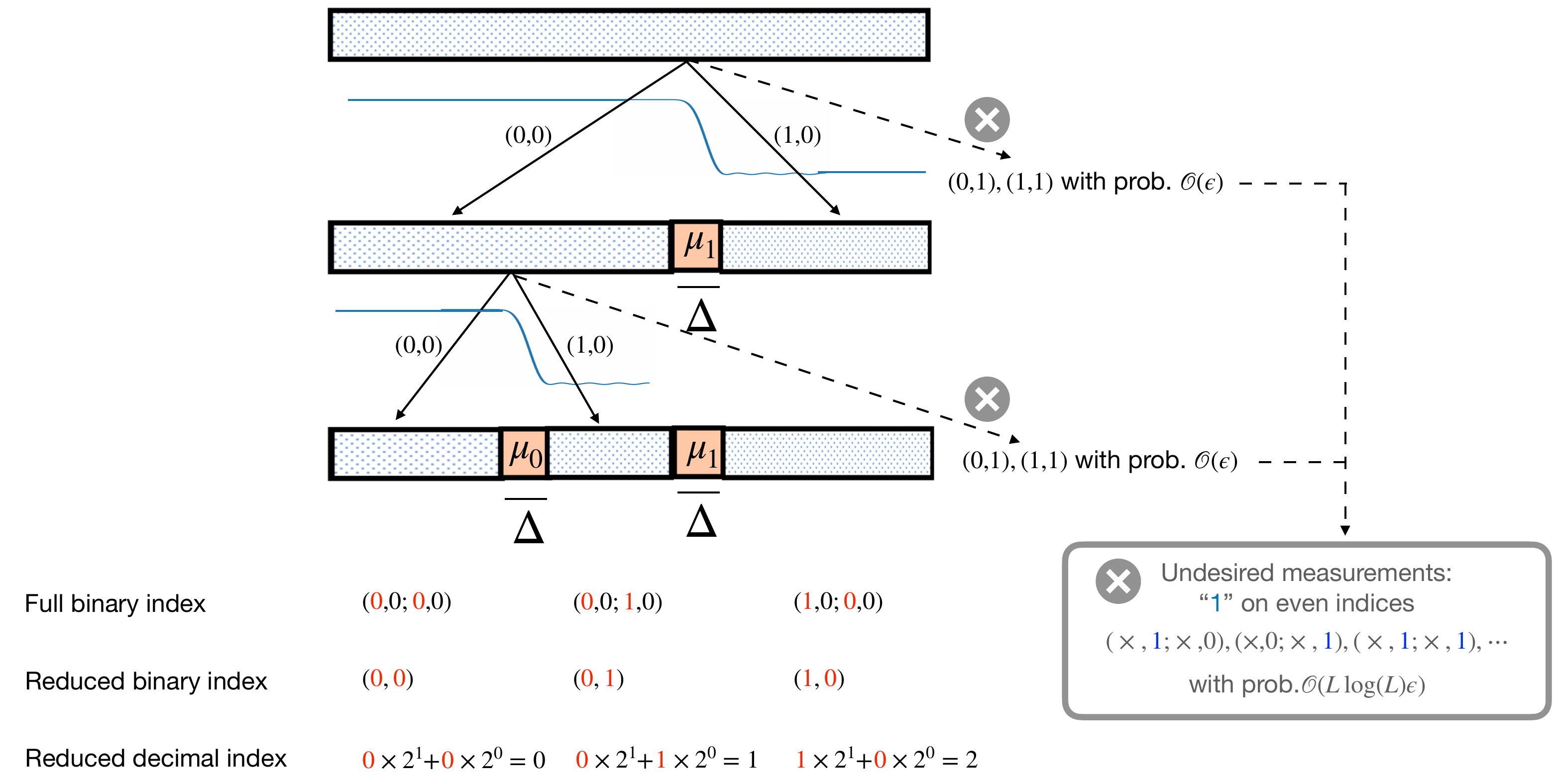}
    \caption{A conceptual visualization of solving multi-band projection problem using \ffqc.}
    \label{fig:feedforward_projector}
\end{figure}

This technical result indicates that using \offqc can implement a quantum channel that is $\Or(\epsilon)$-close to $\varrho \mapsto \Pi_{< \mu} \varrho \Pi_{< \mu} + \Pi_{\ge \mu} \varrho \Pi_{\ge \mu}$. This observation motivates a binary splitting based method for solving the multi-band projection problem. When the measurement outcomes of the monitoring qubit in the previous round indicate the current quantum state supports in a certain energy range, we can adaptively choose a threshold value $\mu$ to apply the projector at the current round. This procedure is exemplified in \cref{fig:feedforward_projector}. It is analogous to traverse a binary tree structure where each leaf stands for the projection onto a certain energy band. By eliminating components sequentially and adaptively, the quantum state collapses to a certain energy band flagged by the sequential measurement outcomes of the monitoring qubit. Consequently, this algorithm  solves the multi-band projection problem using strategic feedforward operations adaptive to intermediate measurement outcomes. This algorithm is given in \cref{alg:feedforward_projector}. Furthermore, the performance of the algorithm is guaranteed through the following theorem, whose proof can be found in \cref{sec:proof-FFQC-projection}.

\begin{thm}[\ffqc for projections]\label{prop:FFQC-projection}
    Suppose the Hamiltonian $H$ has $L$ energy bands with gap at least $\Delta$. There is a \ffqc implementing a quantum channel $\mc{E}_\ffqc$, which is $\epsilon$-close to the exact multi-band projection in trace norm $\norm{\mc{E}_\ffqc - \mc{E}_\mathrm{proj}}_\mathrm{tr} < \epsilon$. To construct that quantum circuit, the number of queries to the block encoding $U_H$ and its inverse is $\wt{\Or}(\norm{H} \Delta^{-1}\log(L) \log(L \epsilon^{-1}))$.
\end{thm}

It is worth noting that the query complexity depends weakly on the number of bands $L$ through polylogarithm. The intuition of that dependency is due to the binary search tree like structure of the algorithm as outlined in \cref{fig:feedforward_projector,alg:feedforward_projector}. This structure also demonstrates the necessities of feedforward since the determination the next binary projection relies on previous measurement outcomes. In the next section, we will compare our \ffqc based methods with other existing methods without using feedforward. We show that the number of queries needed to solve the multi-band projection problem is exponentially worse in $L$ dependency when feedforward is not used in algorithm design.

\begin{breakablealgorithm}
      \caption{Solving multi-band projection problem using \ffqc}
  \label{alg:feedforward_projector}
  \begin{algorithmic}[1]
  \Statex \textbf{Input:} Number of bands: $L$; Spectral parameters: $\mu_0, \cdots, \mu_{L - 1}, \Delta$; Input quantum state $\ket{\phi}$.
    \State Compute $\ell = \lceil \log_2(L)\rceil$.
    \State Initiate $b = (0, 0, \cdots, 0) \in \RR^\ell$ and $s = (0, 0, \cdots, 0) \in \RR^{2 \ell}$.

    \State Denote $\ket{\phi; s} = \ket{\phi}$.
    \For{$j = 1, \cdots, \ell$}
    \State Set $i = \sum_{j = 1}^{\ell} b_j 2^{\ell - j}$.
    \If{$i \ge L$}
    \State \textbf{break} {\color{gray}{ // This handles edge cases when $\log_2(L) \notin \NN$.}}
    \EndIf
   \State Set current index $k = i + 2^{\ell - j}$.
   \State Build a \offqc circuit centering at $\mu_k$ with transition gap $\Delta$.
   \State Input the quantum state $\ket{\phi; s}$, measure two bits $\wt{s}_1, \wt{s}_2$, and collapse onto an output quantum state $\ket{\wt{\phi}}$.
   \State Set $s_{2 j - 1} = \wt{s}_1$ and $s_{2j} = \wt{s}_2$.
   \State Denote $\ket{\phi; s} = \ket{\wt{\phi}}$ (note that $s$ has been updated).
   \State Set $b_j = \wt{s}_1$.
   \EndFor

   \State Set $i = \sum_{j = 1}^{\ell} b_j 2^{\ell - j}$.

    \Statex \textbf{Output:} A quantum state $\ket{\phi; s}$ which supports on the $i$-th band with high probability.
    \end{algorithmic}
\end{breakablealgorithm}

\section{Advantage of feedforward}\label{sec:advantage}

In this section, we discuss the advantage of feedforward by a comparison with existing alternative methods, including a probabilistic projection method, a method based on a random walk on a binary tree, and a method based on adiabatic state preparation. We show that the average query depth of these methods without using feedforwarding depends polynomially on the number of bands $L$. This is exponentially worse than the query depth of our method using feedforwarding in terms of $L$-dependency (see \cref{prop:FFQC-projection}). 

\subsection{Query model for complexity argument}\label{sec:query_model}

We first abstract a query model from our problem setup for the complexity analysis. According to the discussion in the previous section, each cycle of the quantum circuit projects the input quantum state onto an energy subspace flagged by the measurement outcome of ancilla qubits. The problem is to implement a quantum channel represented as $\mc{E}_\mathrm{proj}(\varrho) = \sum_{j \in [L]} \Pi_j \varrho \Pi_j$ in accordance with the projectors onto energy subspaces supported on each band. For any energy subspace $S$, we define the queries in the algorithm design as a pair of projectors $\{\Pi_S, \Pi_{S^\perp}\}$ onto the subspace and its orthogonal complement $S^\perp$, or the ``gabage'' subspace. The actual action among these two projectors is flagged by a binary measurement outcome $s \in \{0, 1\}$. It is worth noting that the energy subspace is not necessarily connected and it can be a union of multiple nonoverlapped components.  This generalizes the definition of the query model, extending it beyond the separation of energy subspaces above or below a specific threshold. We also remark that the binary measurement outcome is for the consideration of preventing undesired shortcuts. When multiple measurement outcomes are allowed, one can design a query $\{\Pi_{S_1}, \Pi_{S_2}, \cdots\}$ simultaneously preparing projectors onto multiple bands. Then, there is a shortcut to solve the target problem that significantly reduces the number of queries. We refer our query model to as \textit{binary projection model}.

\subsection{Method based on random walk on a binary tree}

The feedforward in the algorithm means that the structure of the current cycle depends on the information from previous cycles. For example, in \cref{prop:FFQC-projection}, the choice of the current projector depends on the outcomes of previous measurements. Without the previous measurement outcomes, the strategy in \cref{prop:FFQC-projection} does not work because the bands on which the current state supports are not able to be determined. Consequently, incorrectly determining the current energy range and applying the wrong projector can ruin the results, requiring multiple restarts to solve the problem correctly.

An alternative scheme is to  introduceclassical randomization in the algorithm. For example, we may draw a random number to guess the current energy range and apply corresponding projection strategy. Under binary projection model, applying each projection strategy splits the energy range into two components, which is analogous to the branching of a binary tree. Consequently, the randomized strategy leads to a class of methods which analog random walk on a binary tree. For example, when the current energy range lies in the lower energy subspace, without accessing intermediate measurement results, the algorithm might take a random guess of the current energy range and subsequently misinterprets the current energy range is in the higher energy subspace. Then, by applying the wrong projection strategy due to the wrong guess, the algorithm gets a wrong result and needs to restart. 

In \cref{sec:proof-random-method}, we prove that for any randomized algorithm, the probability that it successfully solves the multi-band projection problem is at most $\Or(L^{-1})$. Hence, to solve the problem with nonvanishing probability, several trials are needed which renders the number of queries at least $\Omega(L)$. This leads to an exponential separation in terms of $L$-dependency between our feedforward method and the randomized method without feedforward. We summarize it as the following theorem and provide its proof in \cref{sec:proof-random-method}.

\begin{thm}\label{thm:binary-projection-lower-bound}
    In the binary projection model,  preparing $\mathcal{E}_\mathrm{proj}$ requires a number of queries to the Hamiltonian at least $\Omega(\log(L))$, where $L$ is the number of bands of the given Hamiltonian. Without using the information of measurement outcomes, the number of queries of methods based on random walk on a binary tree is at least $\Omega(L)$.
\end{thm}

\subsection{Method based on probabilistic projection}\label{sec:prob_projection}

Another method without feedforward is a \textit{probabilistic projection} algorithm. For simplicity, let us assume the input quantum state is a pure state, but the argument can be generalized to any mixed state. Suppose one knows exactly the probability array $\{q(j) := \norm{\Pi_j \ket{\psi}}^2 : j \in [L]\}$. One may choose a band index $j \sim q(\cdot)$ by classical random number generator. Then, the normalized quantum state $\Pi_j \ket{\psi} / \sqrt{q(j)}$ can be implemented by standard coherent quantum algorithms, for example, quantum signal processing or linear combination of unitaries. In terms of the binary projection model, that means keeping only one projector in the queried pair and oblivion the measurement outcome. Taking probabilistic projection, this procedure also implements $\mc{E}_\mathrm{proj}$ by definition. However, to prepare the desired state with high probability, the circuit for projection needs to either classical repetition or amplitude amplification. These contribute to the increase of query depth of each circuit. For a given band index $j$, the amplitude of the desired projected state is $\sqrt{q(j)}$. Consequently, the query depth is either $1 / q(j)$ using classical repetition, or $1 / \sqrt{q(j)}$ using amplitude amplification. Taking the probability of drawing the $j$-th band into account, the average query depth is at least $\sum_{j \in [L]} q(j) \times 1 / \sqrt{q(j)} = \sum_{j \in [L]} \sqrt{q(j)}$. Consider the case where eigenvalues are evenly distributed among bands, namely $q(j) = 1 / L$, the average query depth is at least $\Omega(\sqrt{L})$, which is significantly worse than the logarithmic dependency in the method using feedforward.

\subsection{Method based on adiabatic algorithm}

We compare the feedforward quantum algorithm with another common state preparation approach based on the adiabatic evolution. 
Following the setup in~\cref{sec:prob_projection} on probabilistic projection, we would like to prepare a normalized quantum state that lies in the $j$-th eigenspace of $H$ with known probability $q(j)$. 
The adiabatic approach assumes a time-dependent Hamiltonian $H(t/T)$ over $t \in [0,T]$ such that: (1) the final Hamiltonian $H(1) = H$ is the target Hamiltonian, (2) the initial Hamiltonian $H(0)$ is a simple one with known eigenstates, and (3) the gap condition holds for all $H(t/T), 0 \leq t \leq T$ with the same eigenspace partition as $H$. 
Then, we may first classically sample a band index $j \sim q(\cdot)$, prepare the state $\ket{\psi^{(0)}}$ which lies in the $j$-th band of $H(0)$, and evolve the quantum dynamics governed by the Hamiltonian $H(t/T)$ up to time $T$. 
Then the final state is guaranteed to be approximately the eigenstate of $H(1) = H$ in the $j$-th band.

Notice that the lower bound on the method based on probabilistic projection we discuss in~\cref{sec:prob_projection} does not apply to the adiabatic algorithm. 
This is because, instead of applying projection on the same input state, in the adiabatic approach we start with different input state and evolve it unitarily without a last probability-boosting step.  
However, the adiabatic approach is not efficient in the number of bands $L$ yet. 
Specifically, according to the adiabatic theorem~\cite{JansenRuskaiSeiler2007}, the deepest circuit in the probabilistic adiabatic approach has $\Or(N^{3/2}/L^{3/2})$ depth. 
As long as $L \sim N^{1-\alpha}$ for $0 < \alpha < 1$, which is a necessary condition for the spectral gap to be sublinear in $1/N$, the circuit depth will be $\Or(\mathrm{poly}(L))$, exponentially worse than the feedforward algorithm. 
We refer to~\cref{app:adiabatic} for a more detailed and technical discussion. 

We remark that the $\Or(\mathrm{poly}(L))$ scaling we have shown is an upper bound instead of a lower bound. 
To our knowledge, we are not aware of an adiabatic lower bound for projecting onto an eigenspace with multiple eigenpaths. 
Nevertheless, reference~\cite{BoixoSomma2010} shows that for adiabatically following a single eigenpath, the adiabatic evolution takes at least $\Omega(l_{\mathrm{path}}/\Delta)$ time in the worst case, where $l_{\mathrm{path}}$ is the length of the desired eigenpath. 
As discussed in~\cite{BoixoSomma2010}, the term $l_{\mathrm{path}}/\Delta$ is actually the dominant component in the upper bound of the diabatic error given in~\cite{JansenRuskaiSeiler2007}. 
This implies that the upper bound $\Or(\mathrm{poly}(L))$ we provide based on the adiabatic theorem is likely to coincide with the lower bound in the worst-case example. 
We leave establishing a rigorous lower bound of adiabatic band projection as a future work.

\section{Application in Superconducting Qubits}\label{sec:superconducting}

One natural quantum system that satisfies the problem setting of multi-band projection defined in Sec.~\ref{sec:projection} is superconducting qubits system defined by the Bose Hubbard Hamiltonian. 
A superconducting transmon qubit architecture that allows tunable qubit-qubit coupling \cite{Chen2014} is called the gmon architecture.
To the lowest order of approximation, the system Hamiltonian of gmon qubits consists of  one-body   and  nearest-neighbor-two-body terms represented by  bosonic creation and annihilation operators, $ \hat{a}_j^\dagger $ and $ \hat{a}_j$, and bosonic number operator  $ \hat{n}_j $, for the $ j $-th bosonic mode. In the rotating-wave approximation~(RWA), with a constant rotation rate chosen as the harmonic  frequency of the Josephson junction resonator, the $n$-qubit gmon  Hamiltonian takes the form:
\begin{align}\label{BSHamiltonian} 
H_n(t)= &  \frac{\eta}{2} \sum_{j=1}^n \hat{n}_j(  \hat{n}_j- 1  ) + \sum_{(l,j) \in \mathbb{S}} g_{l,j}(t) ( \hat{a}_l^\dagger\hat{a}_j+\hat{a}_l^\dagger\hat{a}_j) \\\nonumber
&+ \sum_{j=1}^n\delta_j(t) \hat{n}_j  
+\sum_{j=1}^ni f_j(t)\left( \hat{a}_j e^{-   i\varphi_j(t) } -  \hat{a}_j^{\dagger } e^{ i\varphi_j(t) }  \right),
\end{align}
 where the time-constant parameter $ \eta $ represents the anharmonicity of the Josephson junction, and the   time-dependent control parameters are:  (1) amplitude $ f_j(t)$ and (2) phase   $\varphi_j(t) $ of the microwave control pulse; (3)  qubit detuning $\delta_j(t) $,  and (4) tunable capacitive coupling or g-pulse $ g_{l,j}(t) $ between   qubit pair $l, j$ that is connected by a tunable coupler qubit~(defined by edges  $\mathbb{S}$ of qubit coupling graph).
The computational subspace is spanned by the two lowest energy levels  of each   bosonic mode: $ \mathcal{H}_{2 }=\text{Span}\{\ket{0}_j,\ket{ 1}_j\}  $, where $ \ket{n }_j $ represents a Fock state with $ n  $   excitations in   the $ j $-th   mode.  Projected onto the qubit basis, the $n$-qubit Hamiltonian takes the form:
\begin{align}\label{QubitHamiltonian}
H_n(t)= \sum_{(l,j) \in \mathbb{S}} g_{l,j}(t)(X_l X_j + Y_l Y_j ) + \sum_{j=1}^n \left[ \frac{\delta_j(t)}{2} Z_j  - f_j(t) \left(\sin\varphi_j(t) X_j + \cos\varphi_j(t) Y_j \right)   \right].
\end{align}
Each one of the Hamiltonian parameters in Eq.~(\ref{QubitHamiltonian}) can be controlled in real time, with an added Gaussian  error whose standard deviations    are listed in~\cref{Table1}.
\begin{table}[h]
\centering
\begin{tabular}{c @{\hspace{0.3cm}} c@{\hspace{0.3cm}} c@{\hspace{0.3cm}}c@{\hspace{0.3cm}}c@{\hspace{0.3cm}}c@{\hspace{0.3cm}}c } \hline \hline
   &   $g(t)$ & $\delta_j(t)$  & $ f_j(t) $ &  $ \varphi_j(t) $  \\\hline\hline
amplitude  &  $[-20, 20 ]$ MHz& $[-20, 20 ]$ MHz  &  $[-20, 20 ]$ MHz&   $[0, 2\pi]$ \\\hline
error amplitude & $\pm 1$ MHz &$\pm 1$ MHz&$\pm 1$ MHz& $\pm 1$ MHz&     \\
\hline \hline
\end{tabular} 
\caption{Hamiltonian control parameter range.} \label{Table1}
\end{table}

To identify the dominant energy bands, we decompose \cref{BSHamiltonian} into two parts: $H_n(t)=H_{0} + H_{1}(t) $, where $ H_{0}= (\eta/2) \sum_{j=1}^n \hat{n}_j(  \hat{n}_j- 1  )  $ accounts for the large constant-energy gaps separating the energy levels that define qubit subspace from  higher energy levels. Here, the time-dependent Hamiltonian $H_1(t)$ is the controllable terms, and is treated as a perturbative term due to its relatively small magnitude. Consequently, according to the approximation based on perturbation theory,  these determine the minimum energy gap $\Delta\approx \Omega(\eta)$, that separates different energy subspaces that are of the same number of excitations. 

In this way the energy levels of the whole system can be grouped into energy bands labeled by the number of $\ket{2}$, also known as doublons, and the number of $\ket{3}$,  and number of excitations all the way up to inifinite. Energy levels that differ from each other by qubit-level excitations are within the same energy band. For example if the system consists of two qubits, the energy levels are grouped into energy bands as follows
\begin{align}
    &\text{Band} \quad 0 : \{\ket{00},\ket{01},\ket{10},\ket{11}\},\\
   & \text{Band} \quad 1 : \{\ket{02},\ket{20}, \ket{21}, \ket{12} \}\\
   &  \text{Band} \quad 2 : \{\ket{22}\} \\
    &   \text{Band} \quad 3 :  \{\ket{03},\ket{30},\ket{31},\ket{13}\} \\
       \ldots \nonumber
\end{align} Notice that the lowest energy band coincides with the qubit subspace, and projection onto any other subspace indicates information leakage.
Using the algorithm described by \cref{alg:feedforward_projector} allows us to project the Bose-Hubbard system to one of the given subspaces and thus allows us to detect the presence of leakage outside the qubit subspace through sampling. Such a projection has been done locally in \cite{mi2024stable} by local measurement and reset to reduce the entropy of the computation, and is shown to help with state preparation where unwanted excitation leads to leakage outside of qubit subspace. Alternative methods utilize the symmetry properties of different energy bands to post-select away unwanted excitations~\cite{neill2021accurately}. This method, however, cannot be applied in generic band projection problem that doesn't have strictly symmetry properties between different bands, such as ones shown here with Bose Hubbard Hamiltonians.

\section{Conclusion and Outlook}\label{sec:discussion}
In this paper, we have presented a novel framework for quantum algorithm design with QSVT by integrating feedforward operations, resulting in what we term Feedforward Quantum Singular Value Transformation (FQSVT). This advancement leverages intermediate measurements and feedforward controls to utilize quantum information typically discarded in conventional QSVT processes. Our results demonstrate that FQSVT can exponentially accelerate the projection of quantum states onto energy subspaces, outperforming existing methods such as probabilistic projection and adiabatic algorithms in terms of efficiency and query complexity.
We provided detailed theoretical bounds and rigorous proofs to substantiate our findings, underscoring the practical implications of FQSVT in quantum computing and quantum chemistry. Specifically, we illustrated significant applications in managing energy subspaces of superconducting qubits, which are crucial for state preparation and leakage detection. Our framework's ability to efficiently project quantum states onto desired subspaces marks a significant step forward in quantum algorithm design, offering a robust tool for various quantum computing tasks.
The introduction of FQSVT opens several exciting avenues for future research. One promising direction is exploring the connection between FQSVT and quantum error correction codes. Understanding how feedforward operations can be employed to enhance error correction and fault tolerance in quantum systems could lead to more resilient quantum computing architectures. Additionally, further investigations into the underlying correction group structures and more efficient representations of FQSVT circuits could expand its applicability and performance.
Another potential area of research is the integration of FQSVT with other quantum algorithms to address broader classes of problems in quantum simulation and optimization. Using the advantages of feedforward operations, we could develop new quantum algorithms that offer superior performance and scalability.
Moreover, exploring the physical implementation of FQSVT in various quantum hardware platforms, such as trapped ions or photonic systems, could provide valuable insight into the practical challenges and opportunities associated with deploying these advanced quantum circuits. Collaborative efforts between theorists and experimentalists will be crucial in translating the theoretical advancements of FQSVT into tangible improvements in quantum computing technologies.
In summary, FQSVT represents a significant advancement in the field of quantum algorithms, with promising applications and future research directions that could further enhance the capabilities of quantum computing. As we continue to refine and expand this framework, we anticipate that it will play a pivotal role in the development of next-generation quantum technologies.


\newpage
\clearpage
\appendix
\widetext

\section{Quantum singular value transformation circuit and its comprehensive characterization}\label{sec:comprehensive-rep}

This section reviews the relevant theory of Quantum Singular Value Transformation (QSVT). 
 Furthermore, \cref{thm:comprehensive-QSVT,cor:bot-state} present comprehensive characterizations of the matrix representation and the transformed quantum state of QSVT, previously unaddressed in the literature. We will use these findings to prove our technical results in \cref{sec:proof-FFQC-projection}.

QSVT is a powerful quantum algorithm implementing function transformations of general matrices on quantum computers. At the core of QSVT are two pivotal theories: the reduction of high-dimensional matrices to $\mathrm{SU}(2)$ (commonly known as qubitization in standard literature) and an approximation theory within $\mathrm{SU}(2)$, which is referred to as \textit{Quantum Signal Processing (QSP)}. Qubitization establishes a duality between high-dimensional applications and a simpler two-dimensional analytical framework. The approximation theory, meanwhile, simplifies the optimization of complex quantum algorithms to the task of finding the best polynomial approximations for specific target functions. Although the existence of these structures was initially reported in Ref. \cite{LowChuang2017}, their explicit characterization was presented later in Ref. \cite{GilyenSuLowEtAl2019}. In later discovery, Ref. \cite{DongMengWhaleyEtAl2021} demonstrates that a symmetry constraint can be imposed to the parametric ansatz. Furthermore, Ref. \cite{WangDongLin2021} formalizes it as \textit{symmetric QSP} and studies the energy landscape of the related optimization problem. We restate this elegant result concerning approximation using an interleaved parametric $X$ and $Z$ rotations as follows.

\begin{defn}[Quantum signal processing (QSP)]\label{def:QSP}
	Given a set of phase factors $\Psi  = (\psi_0, \cdots, \psi_d) \in \RR^{d+1}$, QSP defines a map from $[-1,1]$ to $\mathrm{SU}(2)$
	\begin{equation}
	x \mapsto U(x, \Psi) := e^{\I \psi_0 Z} \prod_{j=1}^{d} e^{\I \arccos(x) X} e^{\I \psi_j Z}.
	\end{equation}
 A set of phase factors $\Psi = (\psi_0,\psi_1, \cdots, \psi_d)$ is said to be symmetric if $\psi_j = \psi_{d-j}$ for any $j$.
\end{defn}

\begin{thm}[Quantum signal processing \cite{GilyenSuLowEtAl2019,WangDongLin2021}]\label{thm:QSP}
\cref{def:QSP} defines two polynomials $P, Q \in \CC[x]$ through
\begin{equation}\label{eqn:QSP}
        U(x, \Psi) = \left( \begin{array}{cc}
        P(x) & \I \sqrt{1 - x^2} Q(x)\\
        \I \sqrt{1 - x^2} Q^*(x) & P^*(x)
        \end{array} \right).
\end{equation}
Here, $P^*(x)$ and $Q^*(x)$ are polynomials defined by taking complex conjugate on all complex coefficients. These polynomials satisfy
\begin{enumerate}
    \item $\deg(P) \le d$ and $\deg(Q) \le d - 1$,
    \item $P$ has parity $d \mod 2$ and $Q$ has parity $(d-1) \mod 2$,
    \item (Normalization condition) $\abs{P(x)}^2 + (1-x^2) \abs{Q(x)}^2 = 1$ for any $x \in [-1, 1]$.
\end{enumerate}

Conversely, given polynomials $P, Q \in \CC[x]$ satisfying these conditions, there exists a set of phase factors $\Psi$ such that $P, Q$ are parametrized through $U(x, \Psi)$ as in \cref{def:QSP}.

If $\Psi$ is symmetric, then the polynomial $Q \in \RR[x]$ is real. Conversely, given $P \in \CC[x], Q \in \RR[x]$ satisfying these conditions, there exists a set of symmetric phase factors $\Psi$ such that $P, Q$ are parametrized through $U(x, \Psi)$.

\end{thm}

These results can be lifted to high dimensional matrices which is referred to as \emph{quantum singular value transformation} (QSVT). To build the QSVT circuit in \cref{fig:feedforward_mainfig}, the set of phase factors is converted to that adapted to the circuit using the following relation
\begin{equation}\label{eqn:phase-factor-conversion-to-exp}
    \Phi = (\varphi_0, \cdots, \varphi_d) \quad \text{so that} \quad \left\{\begin{array}{l}
             \varphi_0 = \psi_0 - (-1)^d \pi/4,\\
             \varphi_k = \psi_k - (-1)^{d-k} \pi/2\quad \forall k = 1, \cdots, d-1,\\
             \varphi_d = \psi_d - \pi/4.
        \end{array}\right.
\end{equation}

In the realm of quantum computing, the submatrix of a unitary matrix is commonly
referred to as a \textit{block encoding} which is pivotal in representing and manipulating quantum information. Due to the unitary nature of quantum gates, block encoding is also an important input model representing general matrices on a quantum computer.
\begin{defn}[Block encoding] Given an $n$-qubit matrix $H$, if we can find $\alpha, \epsilon \in \mathbb{R}_+$, and an $(m+n)$-qubit unitary matrix $U_H$ so that 
\begin{equation}
\Vert H - \alpha \left(\langle 0^m | \otimes I_N\right) U_H \left( | 0^m \rangle \otimes I_N \right) \Vert \leq \epsilon,
\end{equation}
then $U_H$ is called an $(\alpha, m, \epsilon)$-block-encoding of $H$. 
\label{def:blockencode}
\end{defn}
Specifically, when $\alpha = 1$ (e.g. $\norm{H} \le 1$) and $\epsilon = 0$, the formalism of block encoding means that $H$ coincides with the upper-left submatrix of the unitary block encoding:
\begin{equation}\label{eqn:block-encoding}
    U_H = \begin{pmatrix}
        H & *\\
        * & *
    \end{pmatrix}.
\end{equation}
This special case is referred to as a $m$-block-encoding of $H$, which is considered in the rest of this section for simplicity.

Suppose the singular value decomposition of the matrix is $H = W_1 \Sigma V_1^\dagger$. Following the cosine-sine decomposition of unitary matrices (\cite[Theorem 2.2.2]{Dong2023} and \cite{TangTian2024}), there exists unitary matrices $W_2, V_2 \in \CC^{N(M-1) \times N(M-1)}$ so that 
\begin{equation}\label{eqn:CSD}
    U_H = \begin{pmatrix}
    W_1 & 0\\ 0 & W_2
    \end{pmatrix} \begin{pmatrix}
        \Sigma & S & 0\\ -S & \Sigma & 0\\ 0 & 0 & I_{N(M-2)}
    \end{pmatrix} \begin{pmatrix}
        V_1^\dagger & 0\\ 0 & V_2^\dagger
    \end{pmatrix}.
\end{equation}
Here $S = \sqrt{I_N - \Sigma^2}$ is the supplementary diagonal matrix. The large block diagonal unitary matrices are denoted as
\begin{equation}
    \mathbb{W} := \begin{pmatrix}
        W_1 & 0\\
        0 & W_2
    \end{pmatrix} \text{ and } \mathbb{V} := \begin{pmatrix}
        V_1 & 0\\
        0 & V_2
    \end{pmatrix}.
\end{equation}
We also remark that the matrix in the middle of \cref{eqn:CSD} has a special structure. Let $K$ be a permutation defined as
    \begin{equation*}
        K e_j = e_{2j-1},\ K e_{N+j} = e_{2j}\quad \forall j \in [N],\quad \text{ and } K e_j = e_j \quad \forall 2N \le j < NM
    \end{equation*}
    where $e_j \in \CC^{NM}$ is the standard basis vector whose components are all zero except the $j$-th element that is equal to one. By conjugating with the permutation matrix $K$, this middle matrix can be expressed as a direct sum of $2$-by-$2$ blocks and scalars:
\begin{equation}
\begin{split}
	\left( \begin{array}{ccc}
	\Sigma & S & 0\\
	-S & \Sigma & 0\\
	0 & 0 & I_{N(M-2)}
	\end{array} \right) &= K \bigoplus_{j \in [N]} \left( \begin{array}{cc}
		\sigma_j & \sqrt{1-\sigma_j^2}\\ -\sqrt{1-\sigma_j^2} & \sigma_j
	\end{array} \right) \oplus I_{N(M-2)} K^\dagger\\
	&= K \bigoplus_{j \in [N]} e^{-\I \frac{\pi}{4} Z} e^{\I \arccos(\sigma_j) X} e^{\I \frac{\pi}{4} Z} \oplus I_{N(M-2)} K^\dagger.
\end{split}
\end{equation}
Here, $\sigma_j = \Sigma_{jj}$. Since this rearrangement renders the dimension of each matrix block either one or two, it is termed \textit{qubitization}. When the conjugation of Hadamard gates are ignored, the middle component of the QSVT circuit (see \cref{fig:feedforward_mainfig}(b)) leaves the top ancillary state unchanged. The only difference is that the rotation angles of the controlled rotations are effectively negated when the top ancillary state is switched from $\ket{0}$ to $\ket{1}$. Consequently, the matrix representation of the QSVT circuit is
\begin{equation}
     \mc{Q}(U_H, \Phi) = \frac{1}{2} \begin{pmatrix}
            I_{2NM} & I_{2NM}\\ I_{2NM} & - I_{2NM}
        \end{pmatrix} \begin{pmatrix}
            \mc{U}(U_H, \Phi) & 0\\
            0 & \mc{U}(U_H, -\Phi)
        \end{pmatrix} \begin{pmatrix}
            I_{2NM} & I_{2NM}\\ I_{2NM} & - I_{2NM}
        \end{pmatrix}.
\end{equation}
Here, the middle matrix stands for the interleaving sequence of the controlled rotation and the block encoding, namely,
\begin{equation}
    \mc{U}(U_H, \Phi) := R_\ctrll(\varphi_0) \prod_{k=1}^d \left(U_H^{(-1)^{d-k}} R_\ctrll(\varphi_k)\right).
\end{equation}
with
\begin{equation}
    R_\ctrll(\varphi) = \begin{pmatrix}
        e^{\I \varphi}I_N & 0\\ 0 & e^{-\I \varphi} I_{N(M-1)}
        \end{pmatrix} = K \bigoplus_{j \in [N]} e^{\I \varphi Z} \oplus e^{- \I \varphi} I_{N(M-2)} K^\dagger.
\end{equation}
Note that $\mathbb{W}$ and $\mathbb{V}$ commute with $K$. Also note that the interleaved actions of $U_H$ and $U_H^\dagger$ switch between basis transformations $\mathbb{W}$ and $\mathbb{V}$ depending on the parity of $d$. Consequently, we have
    \begin{equation}\label{eqn:QSVT-mat-rep-1}
    \begin{split}
        & \left.\begin{array}{ll}
            K^\dagger \mathbb{V}^\dagger \mc{U}(U_H, \Phi) \mathbb{V} K &, \text{ when } d \text{ is even},  \\
            K^\dagger \mathbb{W}^\dagger \mc{U}(U_H, \Phi) \mathbb{V} K &, \text{ when } d \text{ is odd},
        \end{array}\right\}\\
        &= \bigoplus_{j \in [N]} e^{\I \varphi_0 Z} \prod_{k=1}^d \left(\begin{pmatrix}
            \sigma_j & \sqrt{1-\sigma_j^2}\\ -\sqrt{1-\sigma_j^2} & \sigma_j
        \end{pmatrix}^{(-1)^{d-k}} e^{\I \varphi_k Z} \right) \oplus e^{- \I \sum_{k=0}^d \varphi_k} I_{N(M-2)}
    \end{split}
    \end{equation}

        Note that
    \begin{equation*}
    \begin{split}
        & \begin{pmatrix}
            \sigma_j & \sqrt{1-\sigma_j^2}\\ -\sqrt{1-\sigma_j^2} & \sigma_j
        \end{pmatrix} = e^{-\I\frac{\pi}{4} Z} e^{\I \arccos(\sigma_j) X} e^{\I\frac{\pi}{4} Z} \\
        \text{ and } & \begin{pmatrix}
            \sigma_j & \sqrt{1-\sigma_j^2}\\ -\sqrt{1-\sigma_j^2} & \sigma_j
        \end{pmatrix}^{-1} = e^{\I\frac{\pi}{4} Z} e^{\I \arccos(\sigma_j) X} e^{-\I\frac{\pi}{4} Z}.
    \end{split}
    \end{equation*}
    Then
    \begin{equation}
    \begin{split}
        &\text{\cref{eqn:QSVT-mat-rep-1}} = \bigoplus_{j \in [N]} e^{\I \varphi_0 Z} \prod_{k=1}^d \left( e^{- \I (-1)^{d-k} \frac{\pi}{4} Z} e^{\I \arccos(\sigma_j) X}  e^{\I (-1)^{d-k} \frac{\pi}{4} Z} e^{\I \varphi_k Z} \right) \oplus e^{- \I \sum_{k=0}^d \varphi_k} I_{N(M-2)}\\
        &= \bigoplus_{j \in [N]} e^{\I (\varphi_0 + (-1)^d \pi/4) Z} \prod_{k=1}^{d-1} \left(e^{\I \arccos(\sigma_j) X}  e^{\I (\varphi_k + (-1)^{d-k}\pi/2) Z}\right) e^{\I \arccos(\sigma_j) X}  e^{\I (\varphi_d + \pi/4) Z}\\
        & \quad\quad\quad \oplus e^{- \I \sum_{k=0}^d \varphi_k} I_{N(M-2)}\\
        &= \bigoplus_{j \in [N]} U(\sigma_j, \iota(\Phi)) \oplus e^{- \I \sum_{k=0}^d \varphi_k} I_{N(M-2)}.
    \end{split}
    \end{equation}
    Here, $\iota : \RR^{d+1} \to \RR^{d+1}$ is a linear map converting the set of phase factors used in constructing the QSVT circuit and to that used in the \text{SU}(2) theory (\cref{thm:QSP}). That is
    \begin{equation}\label{eqn:phase-factor-conversion}
        \Psi = (\psi_0, \cdots, \psi_d) = \iota(\Phi), \quad \left\{\begin{array}{l}
             \psi_0 = \varphi_0 + (-1)^d \pi/4,\\
             \psi_k = \varphi_k + (-1)^{d-k} \pi/2\quad \forall k = 1, \cdots, d-1,\\
             \psi_d = \varphi_d + \pi/4.
        \end{array}\right.
    \end{equation}
    According to \cref{thm:QSP}, $U(x, \iota(\Phi))$ defines two polynomials $P, Q \in \CC[x]$. Let the final basis transformation be
    \begin{equation}\label{eqn:matrix_T}
        \begin{pmatrix}
            T_1 & 0 \\ 0 & T_2
        \end{pmatrix} \text{ with } (T_1, T_2) = \left\{
        \begin{array}{ll}
            (W_1, W_2) & \text{ when } d \text{ is odd}, \\
            (V_1, V_2) & \text{ when } d \text{ is even}.
        \end{array}
        \right.
    \end{equation}
    Then
    \begin{equation}
        \mc{U}(U_H, \Phi) = \begin{pmatrix}
            T_1 & 0\\ 0 & T_2
        \end{pmatrix} \begin{pmatrix}
        P(\Sigma) & \I \sqrt{I_N - \Sigma^2} Q(\Sigma) & 0\\
        \I \sqrt{I_N-\Sigma^2} Q^*(\Sigma) & P^*(\Sigma) & 0\\
        0 & 0 & e^{- \I \sum_{k=0}^d \varphi_k} I_{N(M-2)}
        \end{pmatrix}\begin{pmatrix}
            V_1^\dagger & 0\\ 0 & V_2^\dagger
        \end{pmatrix}
    \end{equation}
    provides a comprehensive matrix representation of the foundational component of QSVT. Specifically, when $H$ is a positive semidefinite Hermitian matrix, we have $H = V \Sigma V^\dagger$ with $W_1 = V_1 = V$ and $\Sigma \succcurlyeq 0$. Then, the matrix representation is simplified as
    \begin{equation}
        \mc{U}(U_H, \Phi) = \begin{pmatrix}
        P(H) & \I \sqrt{I_N - H^2} Q(H) V V_{2, [:M]}^\dagger\\
        \I T_{2,[:M]} V^\dagger \sqrt{I_N-H^2} Q^*(H) & T_2 \left(\begin{array}{cc}
            P^*(\Sigma) & 0 \\ 0 & e^{- \I \sum_{k = 0}^d \varphi_k} I_{N(M-2)}
        \end{array}\right) V_2^\dagger
        \end{pmatrix}
    \end{equation}
    where $T_{2, [:M]}$ and $V_{2, [:M]}$ are matrices derived by extracting the first $M$ columns of $T_2$ and $V_2$ respectively.

    To fully characterize the matrix representation of QSVT circuit, we need to study the lower-right submatrix in which the negated phase factors $- \Phi$ are used. Let $\Psi = \iota(\Phi)$ and $\wt{\Psi} = \iota(-\Phi)$. According to \cref{eqn:phase-factor-conversion}, we have
    \begin{equation*}
        \wt{\psi}_0 = - \psi_0 + (-1)^d \pi/2,\ \wt{\psi}_d = -\psi_d + \pi/2,\ \text{ and } \wt{\psi}_k = -\psi_k + (-1)^{d-k} \pi\quad \forall k = 1, \cdots, d-1.
    \end{equation*}
    Then
    \begin{equation}
        \begin{split}
            U(x,\wt{\Psi}) &= (-1)^d (\I Z) (-1)^{d-1} U(x, -\Psi) (\I Z) = Z U(x,-\Psi) Z = \overline{U(x,\Psi)}\\
            &= \begin{pmatrix}
                P^*(x) & - \I \sqrt{1-x^2} Q^*(x)\\
                -\I \sqrt{1-x^2} Q(x) & P(x)
            \end{pmatrix}.
        \end{split}
    \end{equation}
    As a consequence, when computing $\mc{U}(O_H, -\Phi)$, it suffices to do the substitution $(P, Q) \leftarrow (P^*, - Q^*)$. The full matrix representation of the QSVT circuit is
    \begin{equation}
    \begin{split}
        \mc{Q}(U_H, \Phi) &= \frac{1}{2} \begin{pmatrix}
            I_{2NM} & I_{2NM}\\ I_{2NM} & - I_{2NM}
        \end{pmatrix} \begin{pmatrix}
            \mc{U}(U_H, \Phi) & 0\\
            0 & \mc{U}(U_H, -\Phi)
        \end{pmatrix} \begin{pmatrix}
            I_{2NM} & I_{2NM}\\ I_{2NM} & - I_{2NM}
        \end{pmatrix}\\
        &= \frac{1}{2} \begin{pmatrix}
            \mc{U}(U_H, \Phi) + \mc{U}(U_H, -\Phi) & \mc{U}(U_H, \Phi) - \mc{U}(U_H, -\Phi)\\
            \mc{U}(U_H, \Phi) - \mc{U}(U_H, -\Phi) & \mc{U}(U_H, \Phi) + \mc{U}(U_H, -\Phi)
        \end{pmatrix}.
    \end{split}
    \end{equation}

    When $H$ is a positive semidefinite Hermitian matrix, the matrix components are
    \begin{equation}\label{eqn:sum-of-U}
    \begin{split}
        &\frac{1}{2}\left(\mc{U}(U_H, \Phi) + \mc{U}(U_H, -\Phi)\right)\\
        &= \frac{1}{2} \begin{pmatrix}
            P(H) + P^*(H) & \I \sqrt{I_N-H^2} \left(Q(H) - Q^*(H)\right) V V_{2,[:M]}^\dagger\\
            \I T_{2, [:M]} V^\dagger \sqrt{I_N-H^2} \left(Q^*(H) - Q(H)\right) & A_+
        \end{pmatrix}\\
        &= \begin{pmatrix}
            f(H) & -\sqrt{I_N-H^2} Q_{\Im}(H) V V_{2,[:M]}^\dagger\\
            T_{2, [:M]} V^\dagger \sqrt{I_N-H^2} Q_{\Im}(H) & T_2 \left(\begin{array}{cc}
                f(\Sigma) & 0 \\ 0 & \cos(\sum_{k = 0}^d \varphi_k) I_{N(M-2)}
            \end{array}\right) V_2^\dagger
        \end{pmatrix}
    \end{split}
    \end{equation}
    with
    \begin{equation*}
        A_+ = T_2 \left(\begin{array}{cc}
                P(\Sigma) + P^*(\Sigma) & 0 \\ 0 & (e^{\I \sum_{k = 0}^d \varphi_k} + e^{- \I \sum_{k = 0}^d \varphi_k}) I_{N(M-2)}
            \end{array}\right) V_2^\dagger
    \end{equation*}
    and
    \begin{equation}\label{eqn:diff-of-U}
    \begin{split}
        &\frac{1}{2}\left(\mc{U}(U_H, \Phi) - \mc{U}(U_H, -\Phi)\right)\\
        &= \frac{1}{2} \begin{pmatrix}
            P(H) - P^*(H) & \I \sqrt{I_N-H^2} \left(Q(H) + Q^*(H)\right) V V_{2,[:M]}^\dagger\\
            \I T_{2, [:M]} V^\dagger \sqrt{I_N-H^2} \left(Q^*(H) + Q(H)\right) & A_-
        \end{pmatrix}\\
        &= \begin{pmatrix}
            \I P_{\Im}(H) & \I \sqrt{I_N-H^2} Q_{\Re}(H) V V_{2,[:M]}^\dagger\\
            \I T_{2, [:M]} V^\dagger \sqrt{I_N-H^2} Q_{\Re}(H) & - \I T_2 \left(\begin{array}{cc}
                P_{\Im}(\Sigma) & 0 \\ 0 & \sin(\sum_{k = 0}^d \varphi_k) I_{N(M-2)}
            \end{array}\right) V_2^\dagger
        \end{pmatrix}
    \end{split}
    \end{equation}
    with
    \begin{equation*}
        A_- = T_2 \left(\begin{array}{cc}
                P^*(\Sigma) - P(\Sigma) & 0 \\ 0 & (e^{- \I \sum_{k = 0}^d \varphi_k} - e^{\I \sum_{k = 0}^d \varphi_k}) I_{N(M-2)}
            \end{array}\right) V_2^\dagger.
    \end{equation*}

Using these results, we summarize the full matrix representation of the QSVT circuit as the following theorem.
    
\begin{thm}\label{thm:comprehensive-QSVT}
    Let $H$ be a positive semidefinite Hermitian matrix with $H = V \Sigma V^\dagger$. Let $U_H$ be a $m$-block-encoding. Let $\Phi$ be a set of phase factors and $P, Q \in \CC[x]$ be a pair of polynomials defined by $\Phi$ through \cref{thm:QSP}, and let $f = P_{\Re}$. Suppose $T_2$ and $V_2$ are the unitary matrices defined by the cosine-sine decomposition of the block encoding in \cref{eqn:CSD,eqn:matrix_T}. Then, the matrix representation of the QSVT circuit is as follows:
    \begin{equation}
    \begin{split}
        & \mc{Q}(U_H, \Phi) =\\
        & \begin{psmallmatrix}
            f(H) & - \sqrt{I_N - H^2} Q_{\Im}(H) V V_{2,[:M]}^\dagger & \I P_{\Im}(H) & \I \sqrt{I_N-H^2} Q_{\Re}(H) V V_{2,[:M]}^\dagger\\
            T_{2, [:M]} V^\dagger \sqrt{I_N-H^2} Q_{\Im}(H) & * & \I T_{2, [:M]} V^\dagger \sqrt{I_N-H^2} Q_{\Re}(H) &  A_{1, 3}\\
            \I P_{\Im}(H) & * &  & * \\
            \I T_{2, [:M]} V^\dagger \sqrt{I_N - H^2} Q_{\Re}(H) & * & * & *
        \end{psmallmatrix}
    \end{split}
    \end{equation}
    with 
    \begin{equation}
        A_{1, 3} = - \I T_2 \left(\begin{array}{cc}
                P_{\Im}(\Sigma) & 0 \\ 0 & \sin(\sum_{k = 0}^d \varphi_k) I_{N(M-2)}
            \end{array}\right) V_2^\dagger.
    \end{equation}
\end{thm}
Here, we only explicitly write down submatrices that are used in the proofs in the next section. The submatrices marked with asterisk can be directly found according to \cref{eqn:sum-of-U,eqn:diff-of-U}.

In standard literature, for example Refs. \cite{LowChuang2019,GilyenSuLowEtAl2019}, when acting on an input quantum state $\ket{0^{m + 1}} \ket{\phi}$, the action of the QSVT circuit is denoted as
\begin{equation}
    \mc{Q}(U_H, \Phi)\ket{0^{m+1}}\ket{\phi} = \ket{0^{m+1}} f(H) \ket{\phi} + \ket{\text{``garbage''}}
\end{equation}
with orthogonality $(\bra{0^{m+1}} \otimes I_N) \ket{\text{``garbage''}} = 0$. Such orthogonality condition implies that the ``garbage'' state is eliminated when measuring all $(m+1)$ ancilla qubits with outcome $0^{m+1}$. This quantum state component is discarded in applications by either classical repeated measurements or coherent amplitude amplification. However, in the previous theorem, we reveal the structure of such conventionally considered ``garbage'' state in which the information of the matrix $H$ is transformed through other functions. Such comprehensive characterization enables the reuse of these information and their conversion to usefulness, which will be discussed in the next section. 

Because the ``garbage'' state supports on $(M-1)$ possible ancillary states, directly writing it in the bra-ket notation is convoluted. For notation brevity, we use a mixed vector-ket representation in which a $(NM)$-dimensional vector is block partitioned into $M$ entries with respect to the ancillary basis. Each $N$-dimensional component is a $n$-qubit unnormalized quantum state written in ket. These $M$ entries are ordered according to the decimal value of the ancillary basis $\mathsf{dec}(b_0b_1\cdots b_{m-1}) = \sum_{j\in[m]} b_j 2^{m-1-j}$. Then
\begin{equation}
    \sum_{\vb \in \{0,1\}^m} \ket{\vb} \ket{q_{\vb}} = \begin{pmatrix}
    \ket{q_{00\cdots 00}}\\ \ket{q_{00\cdots 01}} \\ \ket{q_{00\cdots 10}} \\ \vdots \\ \ket{q_{11\cdots 10}} \\ \ket{q_{11\cdots 11}}
    \end{pmatrix}.
\end{equation}
Using this notation, and according to \cref{thm:comprehensive-QSVT}, the ``garbage'' state is explicitly characterized as follows.
\begin{cor}\label{cor:bot-state}
    Suppose the action of the QSVT circuit is
    \begin{equation*}
    \mc{Q}(U_H, \Phi)\ket{0^{m+1}}\ket{\phi} = \ket{0^{m+1}} f(H) \ket{\phi} + \ket{\bot}.
\end{equation*}
Then, under the conditions of \cref{thm:comprehensive-QSVT}, the unnormalized quantum state is
\begin{equation}
    \ket{\bot} = \begin{pmatrix}
        0\\ T_{2, [:M]} V^\dagger \sqrt{I_N-H^2} Q_{\Im}(H) \ket{\phi} \\ \I P_{\Im}(H) \ket{\phi} \\ \I T_{2, [:M]} V^\dagger \sqrt{I_N-H^2} Q_{\Re}(H) \ket{\phi}
    \end{pmatrix}.
\end{equation}
Furthermore, when the QSVT circuit is derived from a set of symmetric phase factors, we have $Q_{\Im} = 0$ and 
\begin{equation}
    \ket{\bot} = \ket{1}\ket{\wt{\bot}} \text{ with } \ket{\wt{\bot}} = \begin{pmatrix}
        \I P_{\Im}(H) \ket{\phi} \\ \I T_{2, [:M]} V^\dagger \sqrt{I_N-H^2} Q_{\Re}(H) \ket{\phi}
    \end{pmatrix} \in \CC^{NM}.
\end{equation}
\end{cor}

\section{Proofs of results in \cref{sec:multi-band-FQSVT}}\label{sec:proof-FFQC-projection}
In this section, we provides proofs for the performance guarantees outlined in \cref{sec:multi-band-FQSVT}. These results are proven based on the comprehensive characterization of the QSVT circuit derived in the previous section.
\begin{proof}[Proof of \cref{lma:1FF-QSVT}]
    When measuring the monitoring qubit, the quantum state collapses to the component entangled with the ancillary qubits state $\ket{0}$ or $\ket{1}$, which is denoted as $\ket{\phi; 0}$ or $\ket{\phi; 1}$ respectively. According to \cref{cor:bot-state}, the unnormalized quantum states after the first measurement are as follows:
    \begin{equation}
        \ket{\phi;0} = \ket{0^m}f(H)\ket{\phi} \text{ and } \ket{\phi;1} = \ket{\wt{\bot}}.
    \end{equation}
    The probability of getting these outcomes in the first measurement is $\bP(s_1) = \norm{\ket{\phi;s_1}}^2$. 

    The circuit construction before applying the second measurement depends on the measurement outcome of the first one. Also note that when $d$ is odd, the second QSVT circuit uses the Hermitian conjugate of the block encoding which differs from the first circuit. That means the second QSVT circuit is $\mc{Q}(U_H^{\mathrm{par}(d)}, \Phi)$ where $\mathrm{par}(d) := (-1)^{d \mod 2}$ be the (signed) parity of $d$. Given the input quantum state $\ket{\psi}$ after the first reset, after acting the second QSVT circuit, the unnormalized quantum state conditioning on that the second measurement on the monitoring qubit gets zero is
    \begin{equation}\label{eqn:state_after_second_meas}
        (\bra{0} \otimes I_{NM}) \mc{Q}(U_H^{\mathrm{par}(d)}, \Phi) \ket{\psi}.
    \end{equation}

    Let us discuss the following cases.
    
        \textbf{Case 1.} The first measurement outcome is equal to $s_1 = 0$. The initialization turns the input quantum state after reset to $\ket{0} \ket{\phi; 0}$. Using the vector-ket notation, we have the unnormalized quantum state conditioning on that the second measurement on the monitoring qubits yields zero is
        \begin{equation}
        \begin{split}
            & \ket{\phi; (0, 0)} := (\bra{0} \otimes I_{NM}) \mc{Q}(U_H^{\mathrm{par}(d)}, \Phi) \ket{0} \ket{\phi; 0} = \begin{pmatrix}
                I_{NM} & 0 
            \end{pmatrix} \mc{Q}(U_H^{\mathrm{par}(d)}, \Phi) \begin{pmatrix}
                f(H) \ket{\phi} \\ 0 \\ 0 \\ 0
            \end{pmatrix}\\
            & = \begin{pmatrix}
                f^2(H) \ket{\phi} \\ 0
            \end{pmatrix} = \ket{0^{m}} f^2(H) \ket{\phi}.
        \end{split}
        \end{equation}

        \textbf{Case 2.} The first measurement outcome is equal to $s_1 = 1$ and $d$ is even. The initialization turns the input quantum state after reset to $\ket{1} \ket{\phi; 1} = \ket{\bot}$. Note that $\mc{Q}(U_H, \Phi)$ is used due to the even $d$. Using the vector-ket notation, we have the unnormalized quantum state conditioning on that the second measurement on the monitoring qubits yields zero is
    \begin{equation}
    \begin{split}
        & \ket{\phi; (1, 0)} := (\bra{0} \otimes I_{NM}) \mc{Q}(U_H, \Phi) \ket{1} \ket{\phi; 1} \\
        & = \begin{psmallmatrix}
            \I P_{\Im}(H) & \I \sqrt{I_N-H^2} Q_{\Re}(H) V V_{2,[:M]}^\dagger\\
            \I V_{2, [:M]} V^\dagger \sqrt{I_N-H^2} Q_{\Re}(H) &  - \I V_2 \left(\begin{array}{cc}
                P_{\Im}(\Sigma) & 0 \\ 0 & \sin(\sum_{k = 0}^d \varphi_k) I_{N(M-2)}
            \end{array}\right) V_2^\dagger
        \end{psmallmatrix} \begin{psmallmatrix}
        \I P_{\Im}(H) \ket{\phi} \\ \I V_{2, [:M]} V^\dagger \sqrt{I_N-H^2} Q_{\Re}(H) \ket{\phi}
    \end{psmallmatrix}\\
    &\stackrel{(1)}{=} \begin{pmatrix}
        (- P_{\Im}^2(H) - (I_N - H^2) Q_{\Re}^2(H) ) \ket{\phi} \\
        \star
    \end{pmatrix}\\
    &\stackrel{(2)}{=} \begin{pmatrix}
        - (I_N - f^2(H)) \ket{\phi} \\
        \star
    \end{pmatrix}.
    \end{split}
    \end{equation}
    The condition that $V_2$ has orthonormal columns is used in equality (1), i.e., $V_{2, [:M]}^\dagger V_{2, [:M]} = I_M$. In equality (2), the normalization condition (\cref{thm:QSP}) is used.
    Here
    \begin{equation}
    \begin{split}
        \star =& \I V_{2, [:M]} V^\dagger \sqrt{I_N-H^2} Q_{\Re}(H) \I P_{\Im}(H) \ket{\phi} \\
        & - \I V_2 \left(\begin{array}{cc}
                P_{\Im}(\Sigma) & 0 \\ 0 & \sin(\sum_{k = 0}^d \varphi_k) I_{N(M-2)}
            \end{array}\right) V_2^\dagger \I V_{2, [:M]} V^\dagger \sqrt{I_N-H^2} Q_{\Re}(H) \ket{\phi}\\
            =& - V_{2, [:M]} V^\dagger \sqrt{I_N-H^2} Q_{\Re}(H) P_{\Im}(H) \ket{\phi} \\
            & + V_2 \left(\begin{array}{cc}
                P_{\Im}(\Sigma) & 0 \\ 0 & \sin(\sum_{k = 0}^d \varphi_k) I_{N(M-2)}
            \end{array}\right) V_2^\dagger V_2 \begin{pmatrix}
                \sqrt{I_N-\Sigma^2} Q_{\Re}(\Sigma) \\ 0
            \end{pmatrix} V^\dagger \ket{\phi}\\
            =& - V_{2, [:M]} V^\dagger \sqrt{I_N-H^2} Q_{\Re}(H) P_{\Im}(H) \ket{\phi} \\
            & + V_2 \underbrace{\left(\begin{array}{cc}
                P_{\Im}(\Sigma) & 0 \\ 0 & \sin(\sum_{k = 0}^d \varphi_k) I_{N(M-2)}
            \end{array}\right) \begin{pmatrix}
                \sqrt{I_N-\Sigma^2} Q_{\Re}(\Sigma) \\ 0
            \end{pmatrix}}_{\sqrt{I_N-\Sigma^2} Q_{\Re}(\Sigma) P_{\Im}(\Sigma)} V^\dagger \ket{\phi}\\
            =& - V_{2, [:M]} V^\dagger \sqrt{I_N-H^2} Q_{\Re}(H) P_{\Im}(H) \ket{\phi} \\
            & + V_{2, [:M]} V^\dagger \sqrt{I_N-H^2} Q_{\Re}(H) P_{\Im}(H) \ket{\phi} \\
            =& 0.
    \end{split}
    \end{equation}
    Consequently, when $d$ is even, we have
    \begin{equation}
        \ket{\phi; (1, 0)} = \begin{pmatrix}
        - (I_N - f^2(H)) \ket{\phi} \\
        0
    \end{pmatrix} = - \ket{0^{m}} (I_N - f^2(H)) \ket{\phi}.
    \end{equation}

    \textbf{Case 3.} The first measurement outcome is equal to $s_1 = 1$ and $d$ is odd. The initialization turns the input quantum state after reset to $\ket{1} \ket{\phi; 1} = \ket{\bot}$. Note that $\mc{Q}(U_H^\dagger, \Phi)$ is used due to the odd $d$. The computation is similar to that in the previous case.
    \begin{equation}
    \begin{split}
        & \ket{\phi; (1, 0)} := (\bra{0} \otimes I_{NM}) \mc{Q}(U_H^\dagger, \Phi) \ket{1} \ket{\phi; 1} \\
        & = \begin{psmallmatrix}
            \I P_{\Im}(H) & \I \sqrt{I_N-H^2} Q_{\Re}(H) V W_{2,[:M]}^\dagger\\
            \I V_{2, [:M]} V^\dagger \sqrt{I_N-H^2} Q_{\Re}(H) &  - \I V_2 \left(\begin{array}{cc}
                P_{\Im}(\Sigma) & 0 \\ 0 & \sin(\sum_{k = 0}^d \varphi_k) I_{N(M-2)}
            \end{array}\right) W_2^\dagger
        \end{psmallmatrix} \begin{psmallmatrix}
        \I P_{\Im}(H) \ket{\phi} \\ \I W_{2, [:M]} V^\dagger \sqrt{I_N-H^2} Q_{\Re}(H) \ket{\phi}
    \end{psmallmatrix}\\
    &\stackrel{(1)}{=} \begin{pmatrix}
        (- P_{\Im}^2(H) - (I_N - H^2) Q_{\Re}^2(H) ) \ket{\phi} \\
        \star
    \end{pmatrix}\\
    &\stackrel{(2)}{=} \begin{pmatrix}
        - (I_N - f^2(H)) \ket{\phi} \\
        \star
    \end{pmatrix}.
    \end{split}
    \end{equation}
    Similarly, the condition $W_{2, [:M]}^\dagger W_{2, [:M]} = I_M$ is used in (1) and the normalization condition is used in (2). Meanwhile, we have
\begin{equation}
    \begin{split}
        \star =& \I V_{2, [:M]} V^\dagger \sqrt{I_N-H^2} Q_{\Re}(H) \I P_{\Im}(H) \ket{\phi} \\
        & - \I V_2 \left(\begin{array}{cc}
                P_{\Im}(\Sigma) & 0 \\ 0 & \sin(\sum_{k = 0}^d \varphi_k) I_{N(M-2)}
            \end{array}\right) W_2^\dagger \I W_{2, [:M]} V^\dagger \sqrt{I_N-H^2} Q_{\Re}(H) \ket{\phi}\\
            =& - V_{2, [:M]} V^\dagger \sqrt{I_N-H^2} Q_{\Re}(H) P_{\Im}(H) \ket{\phi} \\
            & + V_2 \left(\begin{array}{cc}
                P_{\Im}(\Sigma) & 0 \\ 0 & \sin(\sum_{k = 0}^d \varphi_k) I_{N(M-2)}
            \end{array}\right) W_2^\dagger W_2 \begin{pmatrix}
                \sqrt{I_N-\Sigma^2} Q_{\Re}(\Sigma) \\ 0
            \end{pmatrix} V^\dagger \ket{\phi}\\
            =& 0.
    \end{split}
    \end{equation}
    Therefore, when $d$ is odd, the same conclusion also holds
    \begin{equation}
        \ket{\phi; (1, 0)} = - \ket{0^{m}} (I_N - f^2(H)) \ket{\phi}.
    \end{equation}

    \textbf{Remark.} We remark that the use of the $d$-parity dependent circuit $\mc{Q}(U_H^{\mathrm{par}(d)}, \Phi)$ is to cancel the unwanted basis transformation $V_2$ (when $d$ is even) or $W_2$ (when $d$ is odd). This can be seen in the derivations above.

    The probability of measuring these outcomes is $\bP(s_1, s_2) = \norm{\ket{\phi; (s_1, s_2)}}^2$. These prove the theorem.
\end{proof}

When $f$ is a real even polynomial approximating a step function as in \cref{eqn:filter1,eqn:filter2,eqn:filter3}, \cref{lma:1FF-QSVT} implies that
\begin{equation}
    \ket{\phi;(0,0)} \approx \Pi_{\le \mu-\Delta/2}\ket{\phi} \quad \text{and}\quad \ket{\phi;(1,0)} \approx - \Pi_{\ge \mu+\Delta/2}\ket{\phi}.
\end{equation}
That means $\norm{\ket{\phi;(0,0)}}^2 + \norm{\ket{\phi;(1,0)}}^2 \approx 1$, namely $\bP(s_2 = 1) \approx 0$. The rigorous characterization is given in \cref{cor:1FF-QSVT-projection} which is proved as follows.
\begin{proof}[Proof of \cref{cor:1FF-QSVT-projection}]
    Let the index sets of the eigenvalues be
    \begin{equation}\label{eqn:index-set}
        J_{\le \varepsilon} = \{j : \in[N]: E_j \le \varepsilon\} \quad \text{and} \quad J_{\ge \varepsilon} = \{j \in [N]: E_j \ge \varepsilon\}.
    \end{equation}
    Note that $\mathrm{eig}(H) \cap (\mu-\Delta/2,\mu+\Delta/2) = \emptyset$. It implies that $J_{\le}(\mu-\Delta/2) \cup J_{\ge}(\mu+\Delta/2) = [N]$ and any quantum state can be decomposed as
    \begin{equation}\label{eqn:projections}
    \begin{split}
        & \ket{\phi} = \sum_{j \in J_{\le \mu-\Delta/2}} c_j \ket{\phi_j} + \sum_{j \in J_{\ge \mu+\Delta/2}} c_j \ket{\phi_j},\\
        &\Pi_{\le \mu-\Delta/2}\ket{\phi} = \sum_{j \in J_{\le \mu-\Delta/2}} c_j \ket{\phi_j} \text{ and } \Pi_{\ge \mu+\Delta/2}\ket{\phi} = \sum_{j \in J_{\ge \mu+\Delta/2}} c_j \ket{\phi_j}.
    \end{split}
    \end{equation}
    For brevity, we denote $\Pi_{\le} := \Pi_{\le \mu-\Delta/2}$ and $\Pi_{\ge} := \Pi_{\ge \mu+\Delta/2}$ in the rest of the proof. \cref{eqn:filter1,eqn:filter2,eqn:filter3} give the follows. For any $0 \le x \le \mu-\Delta/2$, it holds that
    \begin{equation*}
        \abs{1 -f^2(x)} = \abs{(1+f(x))(1-f(x))} \le 2 \abs{1-f(x)} < \epsilon
    \end{equation*}
    and for any $\mu+\Delta/2 \le x \le 1$, it holds that
    \begin{equation*}
        \abs{f^2(x)} < \epsilon^2/4.
    \end{equation*}
    Then
    \begin{equation}
        \begin{split}
            \norm{\Pi_{\le}\ket{\phi} - \ket{\phi;(0,0)}}^2 &= \norm{\sum_{j\in J_{\le \mu-\Delta/2}} c_j \left(1-f^2(E_j)\right)\ket{\phi_j} + \sum_{j \in J_{\ge \mu+\Delta/2}} c_j f^2(E_j) \ket{\phi_j} }^2\\
            &\le \epsilon^2 \norm{\Pi_{\le}\ket{\phi}}^2 + \left(\frac{\epsilon^2}{4}\right)^2 \norm{\Pi_{\ge}\ket{\phi}}^2 \le \epsilon^2.
        \end{split}
    \end{equation}
    Note that $\ket{\phi; (1,0)} = - \ket{\phi} + \ket{\phi;(0,0)}$, we have
    \begin{equation}
        \norm{-\Pi_{\ge}\ket{\phi} - \ket{\phi;(1,0)}} = \norm{\ket{\phi} -\Pi_{\ge}\ket{\phi} - \ket{\phi;(0,0)}} = \norm{\Pi_{\le}\ket{\phi} - \ket{\phi;(0,0)}} \le \epsilon.
    \end{equation}
    The probability that the second measurement gets one is bounded as
    \begin{equation}
        \begin{split}
            \bP(s_2 = 1) & = 1 - \bP(s_2 = 0) = 1 - \norm{\ket{\phi;(0,0)}}^2 - \norm{\ket{\phi;(1,0)}}^2\\
            &\le 1 - \left(\norm{\Pi_{\le}\ket{\phi}} - \norm{\Pi_{\le}\ket{\phi} - \ket{\phi;(0,0)}}\right)^2 - \left(\norm{\Pi_{\ge}\ket{\phi}} - \norm{\Pi_{\ge}\ket{\phi} - \ket{\phi;(1,0)}}\right)^2\\
            &\le 2\norm{\Pi_{\le}\ket{\phi}}\norm{\Pi_{\le}\ket{\phi} - \ket{\phi;(0,0)}} + 2 \norm{\Pi_{\ge}\ket{\phi}}\norm{\Pi_{\ge}\ket{\phi} - \ket{\phi;(1,0)}}\\
            &\le 2 \sqrt{\norm{\Pi_{\le}\ket{\phi}}^2 + \norm{\Pi_{\ge}\ket{\phi}}^2}\sqrt{\norm{\Pi_{\le}\ket{\phi} - \ket{\phi;(0,0)}}^2 + \norm{\Pi_{\ge}\ket{\phi} - \ket{\phi;(1,0)}}^2}\\
            &\le 2\sqrt{2} \epsilon.
        \end{split}
    \end{equation}
    This completes the proof.
\end{proof}

As a remark, the trace norm of a matrix is defined as
\begin{equation}
    \norm{A}_\mathrm{tr} = \tr{\sqrt{A^\dagger A}}.
\end{equation}

\begin{proof}[Proof of \cref{prop:FFQC-projection}]
    Let us first assume $\log_2(L) := \ell \in \NN$ for simplicity. The analysis can be generalized to any integer value of $L$ with minor modifications. Let $\text{bin} : [L] \to \{0, 1\}^\ell$ be the binary representation operator. Given an integer $i \in [L]$, it is a binary sequence $(\text{bin}_1(i), \cdots, \text{bin}_\ell(i))$ with $\ell$ entries where $\text{bin}_j(i)$ is the $j$-th digit. Conversely, $i = \sum_{j = 1}^{\ell} \text{bin}_j(i) 2^{\ell - j}$. In accordence with \cref{alg:feedforward_projector,fig:feedforward_projector}, we denote the full binary index as an inclusion map $\iota(i) := (\text{bin}_1(i), 0, \text{bin}_2(i), 0, \cdots, \text{bin}_\ell(i), 0)$. Here, the value $0$ at even entry stands for the successful action of the corresponding projection. The full set of measurement results consist of successful and failed implementations, which is $\{0, 1\}^{2 \ell}$. When there is a nonzero even entry, the whole process fails due to the failure of implementing projection. Then, the counter operator is defined as the number of nonzero even entries, namely, $\mathrm{cnt}(s) := \sum_{j = 1}^{\ell} \mathbb{I}_{s^{2j} = 1}$. Consequently, the set of failure results is 
    \begin{equation*}
        \mc{F} := \{0, 1\}^{2\ell} \backslash \{\iota(i) : i \in [L]\} = \cup_{k = 1}^\ell \mc{F}_k \text{ where } \mc{F}_k := \{s : \text{cnt}(s) = k, s \in \mc{F}\}.
    \end{equation*}
    Here, $\mc{F}_k$ stands for the set of failed results consist of $k$ failure counts. 
    
    Following combinatorics, the size of such set is $\abs{\mc{F}_k} = 2^\ell \binom{\ell}{k}$. Furthermore, \cref{lma:1FF-QSVT} implies that the probability of each failed implementation is upper bounded as
    \begin{equation}\label{eqn:bound-prob-Fk}
        \bP(s) \le (2\sqrt{2} \epsilon)^{\mathrm{cnt}(s)}, \forall s \in \mc{F}.
    \end{equation}

    Using these notations, the final density matrix can be factored into two components standing for successful and failed implementations:
    \begin{equation}
        \mc{E}_\ffqc(\varrho_0) = \underbrace{\sum_{i \in [L]} \mc{P}_i \varrho_0 \mc{P}_i^\dagger}_{\text{successful}} + \underbrace{\sum_{s \in \mc{F}} \sigma_s}_{\text{failed}}.
    \end{equation}
    For the first component, the operator $\mc{P}_i \approx \Pi_i$ is the approximate projector onto the $i$-th band prepared by at most $\ell$ \offqc applications. We will show that the first component is close to the exact result $\mc{E}_\mathrm{proj}(\varrho_0)$ because $\mc{P}_i \approx \Pi_i$, and show that the second term is small due to the bounded failure probability in \cref{eqn:bound-prob-Fk}. Applying triangle inequality, we have
    \begin{equation}
        \norm{\mc{E}_\ffqc(\varrho_0) - \mc{E}_\text{proj}(\varrho_0)}_\text{tr} \le \sum_{i \in [L]} \norm{\mc{P}_i \varrho_0 \mc{P}_i^\dagger - \Pi_i \varrho_0 \Pi_i^\dagger}_\text{tr} + \norm{\sum_{s \in \mc{F}} \sigma_s}_\text{tr} 
    \end{equation}
    Note that
    \begin{equation}
    \begin{split}
        \norm{\sum_{s \in \mc{F}} \sigma_s}_\text{tr} & \le \sum_{s \in \mc{F}} \bP(s) \le \sum_{k = 1}^\ell \abs{\mc{F}_k} (2\sqrt{2} \epsilon)^k  = 2^\ell \left( (1 + 2\sqrt{2} \epsilon)^\ell - 1 \right) \\
        & \le 2^\ell \ell 2 \sqrt{2} \epsilon (1 + 2\sqrt{2} \epsilon)^{\ell - 1} = \Or(L \log(L) \epsilon)
    \end{split}
    \end{equation}
    holds when $\epsilon \log(L) = \Or(1)$. This means the second error term is bounded. Telescoping the expression and invoking the inequality $\norm{AB}_\text{tr} \le \norm{A}_2 \norm{B}_\text{tr}$, we have
    \begin{equation}
        \begin{split}
            \norm{\mc{P}_i \varrho_0 \mc{P}_i^\dagger - \Pi_i \varrho_0 \Pi_i^\dagger}_\text{tr} &\le \norm{\mc{P}_i \varrho_0 \mc{P}_i^\dagger - \mc{P}_i \varrho_0 \Pi_i^\dagger + \mc{P}_i \varrho_0 \Pi_i^\dagger - \Pi_i \varrho_0 \Pi_i^\dagger}_\text{tr}\\
            &\le \norm{\mc{P}_i \varrho_0 (\mc{P}_i - \Pi_i)^\dagger}_\text{tr} + \norm{(\mc{P}_i - \Pi_i) \varrho_0 \Pi_i^\dagger}_\text{tr}\\
            &\le \norm{\mc{P}_i - \Pi_i}_2 \norm{\varrho_0}_\text{tr} (\norm{\mc{P}_i}_2 + \norm{\Pi_i}_2)\\
            &\le 2 \norm{\mc{P}_i - \Pi_i}_2
        \end{split}
    \end{equation}
    Note that each $\mc{P}_i$ is a composition of at most $\ell = \log_2(L)$ binary projections. Following \cref{lma:1FF-QSVT}, we have $\norm{\mc{P}_i - \Pi_i}_2 \le \log_2(L) \epsilon$. To conclude, when each binary projection is implemented to precision $\epsilon$, the error in the density matrix is upper bounded as
    \begin{equation}
        \norm{\mc{E}_\ffqc(\varrho_0) - \mc{E}_\text{proj}(\varrho_0)}_\text{tr} \le \Or(L \log(L) \epsilon)
    \end{equation}
    which is independent with the input $\varrho_0$. Consequently, the trace distance of the quantum channel is also bounded by taking the supremum with respect to the input density matrix on both sides.

    Let the target implementation error of the quantum channel be $\hat{\epsilon} = \Or(L \log(L) \epsilon)$. Then, we need to approximate the filter function with error $\epsilon \le \Or(\hat{\epsilon} (L \log(L))^{-1})$ by inverting that relation. Following the complexity of approximating Heaviside function, the number of queries to $U_H$ and $U_H^\dagger$ of each QSVT circuit scales as $d = \Or(\norm{H} \Delta^{-1} \log(\epsilon^{-1}))$. Note that implementing \ffqc in \cref{alg:feedforward_projector} needs at most $2 \lceil \log_2(L) \rceil$ QSVT circuits. Consequently, the total number of queries to $U_H$ and its inverse is
    \begin{equation}
        \Or(d \log(L)) = \wt{\Or}(\norm{H} \Delta^{-1} \log(L) \log(L \hat{\epsilon}^{-1})).
    \end{equation}
    Here, $\wt{\Or}$ hides the $\log\log(L)$ dependency. This completes the proof.
\end{proof}

\section{Proof of \cref{thm:binary-projection-lower-bound}}\label{sec:proof-random-method}

    The preparation of the multi-band projection can be conceptualized as a problem splitting the input quantum state into $L$ bins according to the energy in the eigenbasis. Under the binary projection model (see \cref{sec:query_model} for details), this can be visualized as a binary tree. Each application of binary projection corresponds to a branching on a node on that binary tree. Among the binary projection pair $\{\Pi_S, \Pi_{S^\perp}\}$ for some subspace $S$, we can term the left subtree be the projection onto $S$, while the right subtree be the projection onto $S^\perp$. In order to cover all $L$ bins, the height of the tree should be at least $\lceil \log_2(L) \rceil$ which completes the first half of the statement.

    The major issue ruins the overall complexity is the small probability of success without strategies using previous measurement information. For any given algorithm based on the binary projection model, the algorithm can be conceptually visualized as a decision tree. An algorithm runner may take certain strategy after performing a measurement which projects the quantum state onto some subspace. The strategy takes value in an action space which contains actions including, for example, the correct strategy to be applied, or the wrong strategy leading to the undesired failure of the algorithm. Such a strategy is quantified as a probability $\bP(\text{action} | \tilde{s})$ where the ``state'' $\tilde{s}$ stands for the algorithm runner's knowledge on the outcome of a recent measurement. This is quantified by another confusion probability $\bP(\tilde{s} | s)$ which conditions on the ground truth of the ``state'' $s$. For example, when the full measurement information is known, the confusion probability is degenerated $\bP(\tilde{s} | s) = \delta_{\tilde{s}, s}$. Yet, in the absence of the ability to use feedforward, these two state variables are independent and hence $\bP(\tilde{s} | s) = \bP(\tilde{s})$. Given any algorithm, let $G(\ket{\psi})$ be the probability that the algorithm successfully solves the target multi-band projection problem with an input unnormalized state $\ket{\psi}$. Then, a recurrence of successfully solving the target problem can be established
    \begin{equation*}
    \begin{split}
        G(\ket{\psi}) &\le \bP(\tilde{s} = S | s = S) \bP(s = S) G(\Pi_S \ket{\psi}) + \bP(\tilde{s} = S^\perp | s = S^\perp) \bP(s = S^\perp) G(\Pi_{S^\perp} \ket{\psi})\\
        & = \bP(\tilde{s} = S) \bP(s = S) G(\Pi_S \ket{\psi}) + \bP(\tilde{s} = S^\perp) \bP(s = S^\perp) G(\Pi_{S^\perp} \ket{\psi})\\
        &\le (\bP(\tilde{s} = S) + \bP(\tilde{s} = S^\perp)) \max_{c \in \{S, S^\perp\}} (\bP(s = c) G(\Pi_c \ket{\psi}))\\
        &= \max_{c \in \{S, S^\perp\}} (\bP(s = c) G(\Pi_c \ket{\psi})).
    \end{split}
    \end{equation*}
    Note that each eigenstate component is identically distributed with Haar random input state. Then, the average probability of success can be simplified as a quantity depends on the dimension of the subspace after projection, namely, $\mathcal{G}(L) = \mathbb{E}_\mathrm{Haar}(G(\ket{\psi}))$. Also note that the average probability that the measurement projects the system onto a subspace with Haar random input state is proportional to the dimension of that subspace, namely, $\mathbb{E}_\mathrm{Haar}(\bP(s = c)) = \dim \text{Img}(\Pi_c) / L$. When the subspace $c$ has larger dimension, the probability that the system collapses to that subspace $\bP(s = c)$ is larger. However, due to the larger subspace dimension, successfully solving the target problem is more challenging, namely, $G(\Pi_c \ket{\psi})$ is smaller. Hence, we may argue that the correlation between these two quantities are nonpositive. Therefore, the average probability of success follows the recurrence inequality
    \begin{equation}
        \mc{G}(L) \le \max_{c \in \{S, S^\perp\}} (\mathbb{E}_\mathrm{Haar}(\bP(s = c)) \mathbb{E}_\mathrm{Haar}(G(\Pi_c \ket{\psi}))) \le \max_{k = 1, \cdots, L - 1} \frac{k}{L} \mc{G}(k).
    \end{equation}
    Let us discuss some base cases. When $L = 1$, the state preparation problem is automatically solved, which means $\mc{G}(1) = 1$. When $L = 2$, the bisection of energy bands yield the well classified result without failure, which means $\mc{G}(2) = 1$. Applying induction to the recurrence inequality above, it can be shown that the hypothesis $\mc{G}(L) \le 2 / L$ holds for any $L \ge 2$. Therefore, for any algorithm similar to the previously discussed random binary tree structure without feedforward, the overall probability of success of a single run is at most $2 / L$. In order to ensure a constant probability of success, the algorithm needs to be repeated at least $\Omega(L)$ times, which contributes a multiplicative factor to the overall cost. Hence, the worst-case number of queries without feedforward is at least $\Omega(L)$, which is exponentially worse than the algorithm with feedforward in term of the number of bands $L$.

\section{Adiabatic state preparation}\label{app:adiabatic}

Here we provide more details on the method based on adiabatic algorithm. 
In the probabilistic state preparation problem, our goal is to design a sample strategy which generates an ensemble $\left\{q(j), \ket{\psi_j}\right\}_{j=0}^{L-1}$ such that $q(j)$'s are known probabilities and each $\ket{\psi_j}$ satisfies $\prod_l \ket{\psi_j} = \delta_{lj} \ket{\psi_j}$ ($\delta$ is the Kronecker delta function). 
The adiabatic approach can solve this problem by driving a known state lying in the ``correct'' eigenspaces of a simple Hamiltonian to that of the target Hamiltonian.  
Specifically, let $H^{(0)}$ be another Hamiltonian, which is usually simple with known block encoding and eigenstates. 
Let the interpolating Hamiltonian be 
\begin{equation}
    \widetilde{H}(s) = (1-\gamma(s)) H^{(0)} + \gamma(s) H, \quad 0 \leq s \leq 1
\end{equation}
where $\gamma(s)$ is the scheduling function with $\gamma(0) = 0$ and $\gamma(1) = 1$. 
In the adiabatic approach, we need a stronger gap condition that assumes the same band structure for the interpolating Hamiltonian $\widetilde{H}(s)$ at any time as the final Hamiltonian $H$, and each band path is separated from the rest by a gap $\Delta(s) > 0$. 
The adiabatic state preparation starts from an initial state $\ket{\psi^{(0)}}$ of the ensemble $\left\{q(j), \ket{\psi_j^{(0)}}\right\}_{j=0}^{L-1}$, where $\ket{\psi_j^{(0)}}$ lies in the $j$-th energy band of $H^{(0)}$, and solves the Hamiltonian simulation problem 
\begin{equation}\label{eqn:adiabatic_HSim}
    i \frac{d}{dt} \ket{\widetilde{\psi}(t)} = \widetilde{H}(t/T) \ket{\widetilde{\psi}(t)}, \quad 0 \leq t \leq T, \quad \ket{\widetilde{\psi}(0)} = \ket{\psi^{(0)}}. 
\end{equation}
Then the final state is approximately a sample from $\left\{q(j), \ket{\psi_j}\right\}_{j=0}^{L-1}$ as desired for sufficiently long evolution time $T$. 

The adiabatic theorem bounds the distance between the state $\ket{\widetilde{\psi}(T)}$ and the ideal eigenspaces. 
Here we state the result from~\cite{JansenRuskaiSeiler2007}. 

\begin{lem}[{Adiabatic theorem~\cite[Theorem 3]{JansenRuskaiSeiler2007}}]\label{lem:adiabatic_theorem}
    Suppose that the interpolating Hamiltonian $\widetilde{H}(s)$ is twice continuously differentiable, and the adiabatic dynamics starts from a state $\ket{\psi_j^{(0)}}$ in the $j$-th energy band of $H^{(0)}$. 
    Let $M_j$ be the number of eigenpaths of $\widetilde{H}(s)$ in the $j$-th energy band. 
    Then 
    \begin{equation}
    \begin{split}
        & \quad \norm{ \Pi_{j} \ket{\widetilde{\psi}(T)} - \ket{\widetilde{\psi}(T)} } \\
        & \leq \frac{1}{T} \left( \frac{M_j \norm{ \widetilde{H}'(0) }}{\widetilde{\Delta}(0)^2} + \frac{M_j \norm{ \widetilde{H}'(1) }}{\widetilde{\Delta}(1)^2} + \int_0^1  \left( \frac{M_j \norm{ \widetilde{H}''(s) }}{\widetilde{\Delta}(s)^2} + \frac{7M_j\sqrt{M_j} \norm{ \widetilde{H}'(s) }^2}{\widetilde{\Delta}(s)^3} \right) d s \right). 
    \end{split}
    \end{equation}
\end{lem}

According to~\cref{lem:adiabatic_theorem}, we can increase the evolution time $T$ to reduce the leakage rate out of the desired eigenspace. 
In a simplified scenario where we assume $\norm{H^{(0)}} = \Or(\norm{H}) = \Or(1)$ and choose the scheduling function with uniformly bounded derivatives $|\gamma'|, |\gamma''| = \Or(1)$, it suffices to choose $T = \Or(M_j^{3/2}  / (\epsilon \min\widetilde{\Delta}^{3}) ) $ to bound the diabatic error out of the $j$-th energy band by $\Or(\epsilon)$. 
If we would like to bound the diabatic error for each sample of the initial state, then it suffices to choose $T = \Or(M^{3/2}  / (\epsilon \min\widetilde{\Delta}^{3}) ) $ where $M = \max_j M_j$ is the maximum number of eigenpaths in one energy band. 

To construct a digital quantum algorithm, we can apply the truncated Dyson series method~\cite{LowWiebe2019} to solve the Hamiltonian simulation problem in~\cref{eqn:adiabatic_HSim}. 
It requires $\widetilde{\Or}( \norm{H} T \log(1/\epsilon) )$ queries to the block-encoding of $\widetilde{H}(s)$, which can be constructed with one query to the controlled version of the block-encodings of $H^{(0)}$ and $H$. 
Putting everything together, we obtain a complexity estimate of the adiabatic state preparation approach. 

\begin{prop}\label{prop:adiabatic_state_prep}
    Let $H$ be a Hamiltonian with $\norm{H} = 1$. 
    Let $M$ be the maximum number of eigenpaths in one energy band of $\widetilde{H}(s)$, and $\widetilde{\Delta}(s)$ be its spectral gap between different bands. 
    Suppose that we use the initial Hamiltonian $H^{(0)}$ with $\norm{H^{(0)}} = \Or(1)$ and choose the scheduling function $\gamma(s)$ with $|\gamma'(s)|, |\gamma''(s)| = \Or(1)$. 
    Then the adiabatic approach can sample from an ensemble $\left\{ p_l, \ket{\psi_l} \right\}$ such that $\norm{\prod_l \ket{\psi_l} - \ket{\psi_l} } \leq \epsilon$, using 
    \begin{equation}
        \widetilde{\Or}\left( \frac{M^{3/2} }{\epsilon (\min\widetilde{\Delta}(s))^3 } \right) 
    \end{equation}
    queries to the block-encoding of $H$. 
\end{prop}

According to~\cref{prop:adiabatic_state_prep}, the adiabatic state preparation approach is not efficient if an energy band of interest has too many eigenvalues. 
Specifically, the overall complexity has a polynomial dependence on the number of eigenvalues in that energy band. 
This polynomial dependence comes from the adiabatic theorem in~\cite{JansenRuskaiSeiler2007}, which is by far the best result for bounding the adiabatic errors in the case with multiple eigenpaths and not clear whether further improvement is possible or not. 
Notice that~\cref{prop:adiabatic_state_prep} implies that adiabatic approach will always have a $\Or(\text{poly}(N))$ dependence: if $L = \Or(N^{1-\alpha})$ for $\alpha > 0$, then $M \geq N/L \geq \Omega(N^{\alpha})$ so the adiabatic query complexity is polynomial in $N$ (and also in $L$), and if the number of energy bands $L = \wt{\Theta}(N)$, then, although in this case $M = \wt{\Or}(1)$, the spectral gap of $H$ is at most $\wt{\Or}(1/N)$ so the overall query complexity is still polynomial in $N$. 

\cref{prop:adiabatic_state_prep} also implies a cubic dependence on the inverse spectral gap and a linear dependence on the inverse error, but these two scalings are improvable. 
For the gap dependence, if we know the simultaneous information of the gap $\widetilde{\Delta}(s)$ for all $s$, then we may reduce the gap dependence to be linear by carefully choosing the scheduling function and only slowing it down when the gap is small. 
Such a strategy has been successfully applied to construct optimal quantum algorithms for unstructured search problem~\cite{RolandCerf2002,AlbashLidar2018} and linear system problem~\cite{AnLin2019}. 
For the error dependence, it can also be reduced to a poly-logarithmic scaling by using the scheduling function with boundary cancellation condition~\cite{Nenciu1993,GeMolnarCirac2016}.

\section{Defining general feedforward quantum algorithms}\label{sec:general-feedforward}
In the main text, we detail a framework for integrating intermediate measurements and feedforward operations with QSVT. The concept of feedforwarding extends to broader quantum algorithm structures. This section presents a general definition of feedforward quantum algorithms.

A quantum algorithm can be considered as a quantum circuit with a tunable parametric structure, which can be, for example, variable quantum-gate parameters or circuit architectures. Consequently, we may identify the set of all potential candidate quantum circuit representations of quantum algorithms as a set $\Theta$ which is referred to as a \textit{parameter space}. Given a feasible parameter $\theta \in \Theta$, the corresponding quantum algorithm is denoted as $\mc{U}(\theta)$ where $\mc{U}$ is a \textit{circuit instruction} which maps the gate parameters to a quantum circuit with a sequential actions of gates. To extract information from the system, we also need to perform measurements and reset measured qubit state for the computation in the next round. This process is denoted as a measurement-and-reset (MAR) operation $\mc{M}$ which is a probabilistic mapping of quantum state with measurement outcome. For example, when measuring the top ancilla qubit, the MAR operation can be
\begin{equation}
    \mc{M}(\ket{0}\ket{\phi_0} + \ket{1}\ket{\phi_1}) = \left\{
    \begin{array}{ll}
        \ket{0} \ket{\phi_0} / \norm{\ket{\phi_0}} & \text{ and output } 0 \text{ with probability }\norm{\ket{\phi_0}}^2, \\
        \ket{1} \ket{\phi_1} / \norm{\ket{\phi_1}} & \text{ and output } 1 \text{ with probability }\norm{\ket{\phi_1}}^2.
    \end{array}
    \right.
\end{equation}
In some applications, the quantum state after measurement collapse is not relevant and only the measurement outcomes are used to design the protocol in the next round. Then, the MAR operation outputs the measurement outcome and resets the quantum state to a default one for the next round. For example, estimating energy and gradients in variational quantum eigesolver \cite{PeruzzoJarrodYungEtAl2014} exemplifies this application. 

Let us consider the measurement outcome of the $i$-th round be $s_i$. Then, the circuit structure of the next round $\theta_{i + 1}$ is determined by a feedforward protocol based on the record of measurement outcomes $\mc{S}_i = (s_0, s_1, \cdots, s_i)$ and previous circuit structures $\mc{A}_i = (\theta_0, \theta_1, \cdots, \theta_i)$. Consequently, the determination of the circuit structure of the next round can be conceptualized as a \textit{policy function} $\mc{P}$ that is either deterministic $\theta_{i + 1} = \mc{P}(\mc{A}_i, \mc{S}_i)$ or random $\theta_{i + 1} \sim \mc{P}(\cdot | \mc{A}_i, \mc{S}_i)$. 

With an initial circuit parameter $\theta_0$, the design of the quantum algorithm is to find a candidate quantum algorithm in the parameter space $\Theta$ so that a desired problem is solved accurately with reasonable cost. This algorithmic design can be solved using either explicit derivation or optimization-based search. The success of solving the problem is measured by a cost metric $\mc{C}$. For example, in a quantum control problem in pulse design \cite{NiuBoixoSmelyanskiyEtAl2019}, it can be
\begin{equation*}
    \mc{C} = \mathrm{weighted\_combination}\left\{ \mathrm{target\_infidelity}, \mathrm{gate\_prameter\_cost}, \mathrm{circuit\_depth}, \cdots \right\}.
\end{equation*}

\begin{figure}[htbp]
    \centering
    \includegraphics[width=\linewidth]{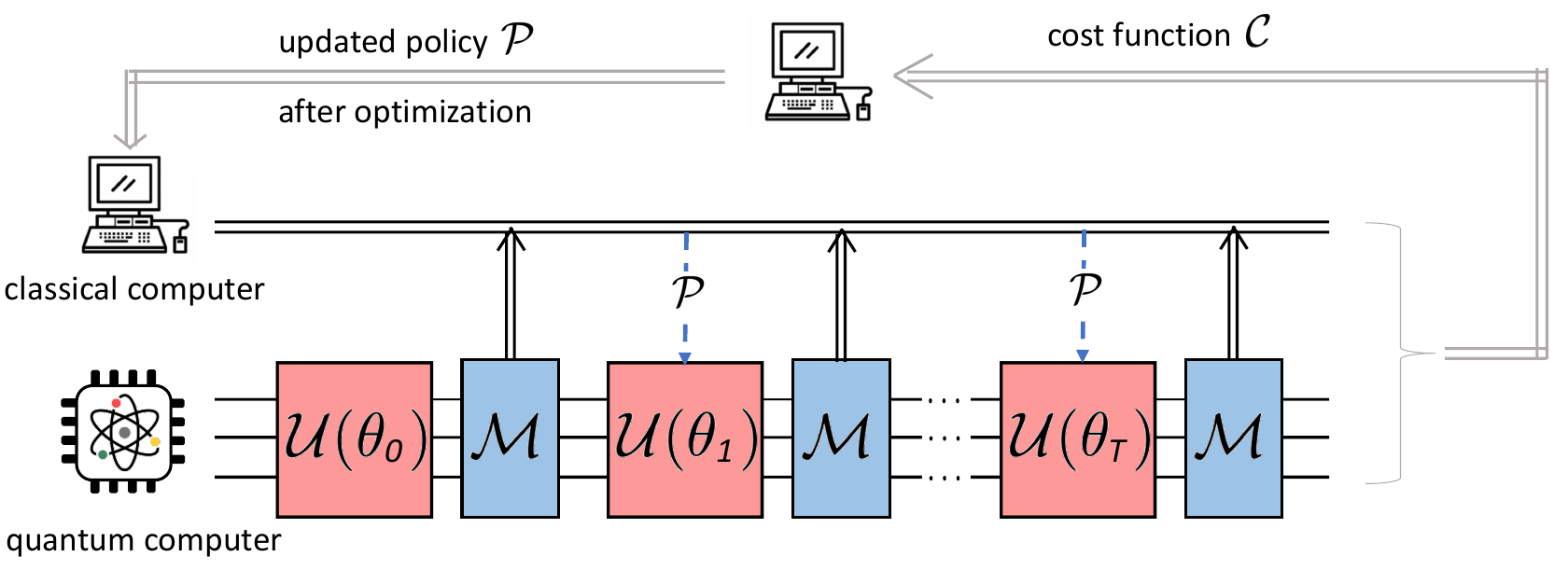}
    \caption{A framework of feedforward quantum algorithms.}
    \label{fig:feedforward_general}
\end{figure}

Putting them together gives a general definition of feedforward quantum algorithms. We also visualize the workflow of feedforward quantum algorithms in \cref{fig:feedforward_general}.

\begin{defn}
    A feedforward quantum algorithm is a tuple $(\Theta, \mc{U}, \mc{M}, \mc{P}, \mc{C}, \theta_0)$. From an initial circuit structure parameter $\theta_0$, the circuit structure takes values in the parameter space $\Theta$ based on the record of measurement outcomes and previous circuit structures. The design policy $\mc{P}$ is derived by minimizing the cost metric $\mc{C}$.
\end{defn}

\end{document}